\documentclass[preprint,11pt,number]{elsarticle}

\usepackage[margin=1in]{geometry}
\usepackage{amssymb}
\usepackage{amsmath}
\usepackage{natbib}
\usepackage{float}
\usepackage{booktabs}
\usepackage{multirow}
\usepackage{siunitx}
\usepackage{array}
\usepackage{longtable}
\usepackage{makecell}
\usepackage{tabularx}
\usepackage{rotating}
\usepackage{lineno}
\usepackage{tabularx}
\usepackage[T1]{fontenc}
\usepackage[utf8]{inputenc}
\usepackage{calc}
\usepackage{indentfirst}
\usepackage{fancyhdr}
\usepackage{graphicx,epstopdf}
\usepackage{lastpage}
\usepackage{ifthen}
\usepackage{setspace}
\usepackage{enumitem}
\usepackage{mathpazo}
\usepackage{titlesec}
\usepackage{etoolbox} 
\usepackage{tabto} 
\usepackage{xcolor, colortbl} 
\usepackage{soul} 
\usepackage{microtype} 
\usepackage{tikz} 
\usepackage{refcount} 
\usepackage{changepage} 
\usepackage{attrib} 
\usepackage{upgreek} 
\usepackage{pbox} 
\usepackage{ragged2e} 
\usepackage[subfigure]{tocloft} 
\usepackage{marginnote} 
\usepackage{marginfix} 
\usepackage{enotez} 
\usepackage{xstring} 
\usepackage[labelformat=simple]{subfig} 
 
\usepackage[skip=0.5\baselineskip]{caption} 

\journal{Journal of Hydrology: Regional Studies}

\begin{document}

\begin{frontmatter}



\title{Bias correction of satellite and reanalysis products for daily rainfall occurrence and intensity} 


\def\correspondingauthor{\footnote{Corresponding author.
		E-mail address: john.bagiliko@aims-senegal.org}}

\author[1,2]{John Bagiliko \correspondingauthor{}} 
\author[3]{David Stern}
\author[4]{Francis Feehi Torgbor}
\author[1,2,3]{Danny Parsons}
\author[5]{Samuel Owusu Ansah}
\author[1]{Denis Ndanguza}

\affiliation[1]{organization={Department of Mathematics, School of Science, College of Science and Technology, University of Rwanda },
            addressline={P.O. Box 3900}, 
            city={Kigali},
            country={Rwanda}}
        
\affiliation[2]{organization={African Institute for Mathematical Sciences, Research and Innovation Center},
	addressline={Rue KG590 ST}, 
	city={Kigali},
	country={Rwanda}}

\affiliation[3]{organization={IDEMS International},
	addressline={RG2 7AX}, 
	city={Reading},
	country={United Kingdom}}

\affiliation[4]{organization={Ghana Innovations in Development, Education and the Mathematical Sciences},
	addressline={Okponglo, East Legon}, 
	city={Accra},
	country={Ghana}}

\affiliation[5]{organization={Ghana Meteorological Agency},
	addressline={ P.O. Box LG 87}, 
	city={Accra},
	country={Ghana}}

\begin{abstract}
In data-sparse regions with limited rain gauge coverage, satellite and reanalysis rainfall estimates (SREs) are a vital complementary data source. However, their utility is fundamentally limited by inherent biases, necessitating correction before use. This study evaluates a comprehensive suite of bias correction (BC) methods, including traditional statistical approaches (LOCI, QM) and machine learning methods (SVR, GPR), applied to seven SREs across 38 stations in Ghana and Zambia. Notably, the LOCI method was implemented using a constrained version introduced in this work to prevent the generation of unrealistically high rainfall values observed with the original approach. Results indicate that statistical methods (QM and LOCI) generally outperformed the machine learning techniques, although QM exhibited a known tendency to inflate rainfall values. SREs corrected with the statistical methods demonstrated a high capability for detecting dry days across all stations (POD $\ge 0.80$). The ENACTS product, which uniquely integrates a large number of station records, emerged as the most amenable to corrections in Zambia, nearly all BC methods successfully reduced the mean error in daily rainfall amounts at over 70\% of stations. However, its performance was not consistently high at Moorings, a station not incorporated into the ENACTS product. This indicates that the product’s performance may be less reliable at independent stations, highlighting the need for further validation at additional independent locations. A critical limitation persisted, however, as nearly all SREs (except ENACTS), even after correction, consistently failed to improve the detection of heavy and violent rainfall events (POD $\le 0.2$). This limits the utility of these corrected SREs for applications such as flood risk assessment, highlighting a crucial research gap for future research focusing specifically on these events.
\end{abstract}


\end{frontmatter}


\section{Introduction}\label{sec1}

In data-sparse regions with limited rain gauge coverage, SREs serve as an important
complementary data source. These products deliver precipitation information at fine spatial and temporal resolutions, often with near-global coverage~\cite{Feidas2009, Monsieurs2018} and spatial resolutions as high as 0.05°~\cite{Funk2014, Funk2015}.

However, SREs are known to be associated with different shortcomings, such as overestimation of rain day frequency~\cite{MAPHUGWI2024107718}, overestimation or underestimation of rainfall intensities at different geographical areas, and challenge in detecting extreme rainfall events at various locations~\cite{Ageet2022, Mekonnen2023}. These shortcomings could be due to indirect measurement techniques, dependence on secondary variables, and potential flaws in temporal coverage, spatial resolution, retrieval algorithms, or sensor precision, reducing their dependability for climate-related analyses~\cite{Dinku2018, Tot2015, amt-11-1921-2018}. Consequently, bias correction (BC) remains an essential step to enhance the utility of SREs for climate applications~\cite{Okirya2025}.

Numerous studies have explored traditional statistical BC techniques to improve SRE accuracy~\cite{ZHANG2022101192, 10.2166/wcc.2020.261, ABERA2016471, Schmidli2006, Gudmundsson2012, ATIAH2023e17604, Nigussie2022, KatiraieBoroujerdy2020, Mendez2020, LAKEW2020100741, Shen2021, Chaudhary2019, DEHAGHANI2023101360, LOBER2023101386, ASILEVI2024101610}. Traditional methods fall into two broad categories: mean-based, such as Local Intensity Scaling (LOCI), which adjust long-term averages to match gauge observations, and distribution-based methods, such as Quantile Mapping (QM), which align the statistical distributions of SREs with ground data~\cite{Soo2019}. Many works have found QM and other distribution-based method to outperform other traditional statistical methods under multiple contexts for precipitation correction~\cite{cli12120226, LAKEW2020100741}. However, a known limitation of this method is its tendency to inflate values~\cite{Maraun2013}. Some recent studies have leveraged ML algorithms~\cite{Tao2016, Chen2010, Li2023, BAEZVILLANUEVA2020111606, Le2023, TRIPATHI2006621}, which all demonstrate a potential for reducing biases in SREs. Some methods specifically correct both rain day frequency and rainfall intensity. For example, LOCI and QM adjust both rain or no rain days and rainfall intensities~\cite{Maraun2013}. In the context of bias correction, the ML approaches are generally regression-based, or a combination of classification and regression. The regression-based ML methods only correct the rainfall intensity. Others attempt to correct both rain day frequency and rainfall intensity with ML. For example, 
~\citet{Chen2010} demonstrated an approach where numerous meteorological variables were used to first develop a rain or no-rain classifier, followed by regression models applied exclusively to wet days. Based on the notion that satellite precipitation inversion errors vary with intensity, ~\citet{Li2023} developed regression models (specifically Gaussian Process Regression) for different rainfall intensity classes (i.e., 0--0.1, 0.1--10, 10--25, 25--50, 50--100, and >100~mm/day) which demonstrated good performance. 

Performance of BC methods seem to depend on SRE, temporal scale, and location. ~\citet{DHAWAN2024e40352} compared statistical methods (Linear scaling, Variance scaling, Power transformation, QM, Multivariate bias correction) and ML methods (Decision tree regressor, Extreme gradient boosting, Support Vector Regression (SVR)) for bias-correcting ERA5-Land in two provinces of Italy, and found that the statistical methods outperformed the ML methods across monthly and daily time scales, while the ML methods had the best performance on hourly time scale. They also noted the performance of their BC methods declined from coarser temporal scales (monthly) to finer, more granular temporal scales (hourly). 

Despite these efforts, comparative evaluations of BC methods remain limited for Africa (including our study area), particularly for newer SREs such as the Precipitation Estimation from Remotely Sensed Information using Artificial Neural Networks-Climate Data Record (PERSIANN-CDR, denoted PCDR)~\cite{Ashouri2015}, and Enhancing National Climate Services (ENACTS)~\cite{Dinku2017, Dinku2022}. This gap is critical, given the continent’s diverse climatic regimes and high reliance on rainfall for agriculture. 

It is worth noting that the ENACTS rainfall product is distinct in its methodology, directly merging extensive daily station observations (over 40 stations) from Zambia with bias-corrected CHIRPS (Climate Hazards Group Infrared Precipitation with Stations) or TAMSAT (Tropical Application of Meteorology using Satellite data) satellite estimates~\cite{Dinku2022}. This contrasts with CHIRPS, which incorporates station data only on a 5-day basis without full historical integration~\cite{Funk2015}, and TAMSAT, which uses station data purely for calibration rather than direct merging~\cite{Maidment2017}.   

This study evaluated the performance of four BC methods (LOCI, QM, SVR and GPR)
applied to seven SREs across 38 stations in Ghana and Zambia. The uncorrected SREs were also validated alongside the corrected estimates. The BC methods were chosen to include mean-, distribution-, and ML-based approaches. It is important to note that the LOCI used here is a constrained version introduced in this work, as we identified a potential for the original version to generate some unrealistically high values when applied in our study area. This constrained LOCI is detailed in Section~\ref{loci_sec}. For all our ML methods, we built separate models for different precipitation intensity classes (except days which SREs estimate as dry (i.e., <0.85 mm/day)). The study region covers three major African rainfall regimes~\cite{bagiliko2025}, enabling a robust assessment of method efficacy under varying climatic conditions. Statistical evaluation metrics employed include Mean Error (ME), Correlation Coefficient (Corr), Ratio of Standard Deviations (RSD), Root Mean Squared Error (RMSE), Mean Absolute Error (MAE), the Nash--Sutcliffe efficiency (NSE), and Probability of Detection (POD) aimed at assessing various aspects of the corrected estimates with respect to the gauge data as well as the uncorrected SRE estimates.

This paper is arranged as follows: Section~\ref{sec2} describes the study area and data, presents the BC methods as well as the evaluation metrics; Section~\ref{sec3} presents the results; Section~\ref{sec4} presents the discussion; the conclusion is given in Section~\ref{sec5} while the limitations of the study are presented in Section~\ref{sec6}.

\section{Materials and methods}\label{sec2}
\subsection{Study area and data}\label{sec_material}
\begin{figure}[H]
	\centering
	\includegraphics[width=.6\linewidth]{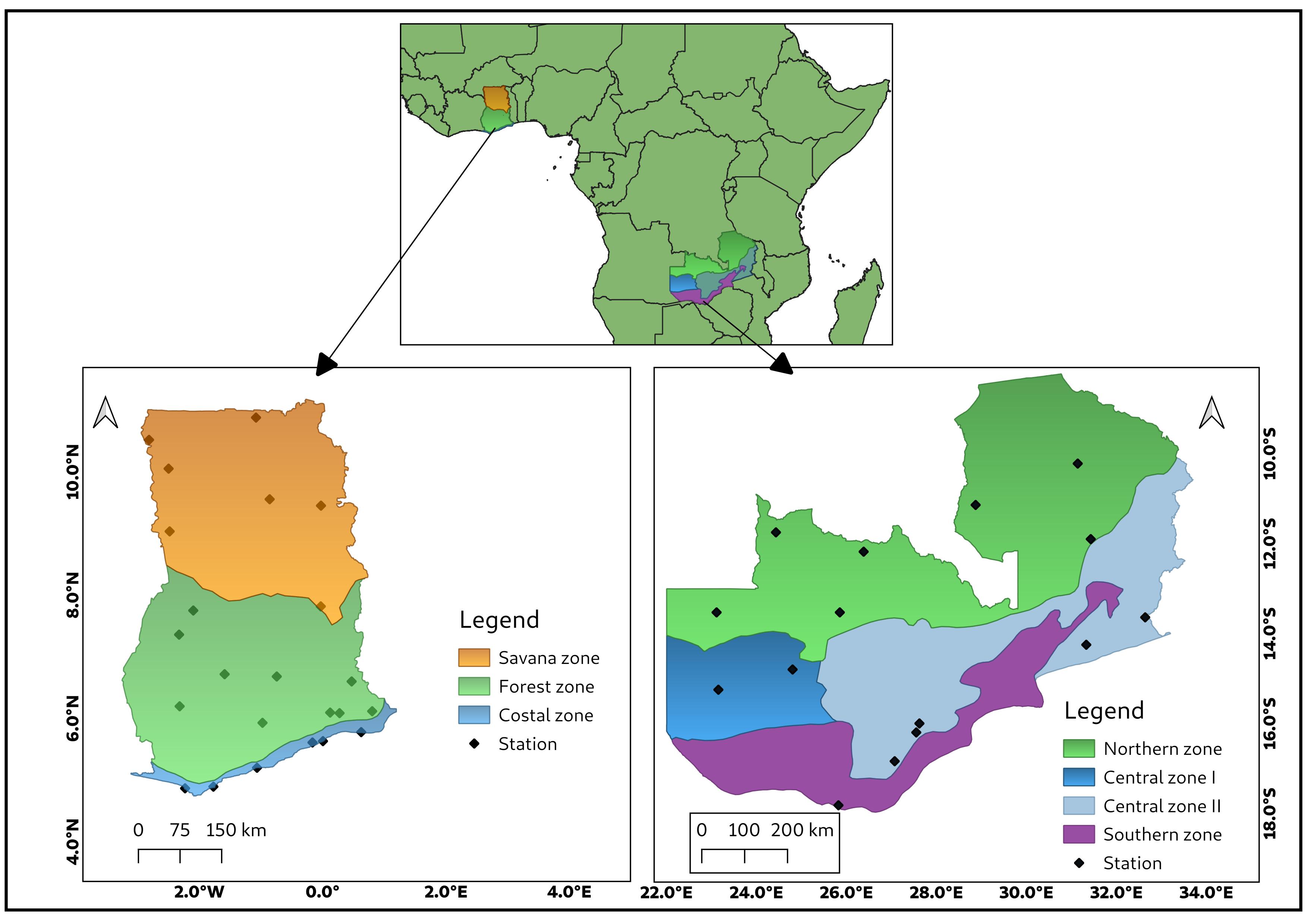}
	\caption{Map of the study area showing Ghana's climatic zones (bottom left) adapted from \cite{Bessah2022}, and Zambia's agroecological zones (bottom right)} \label{fig_study_area}
\end{figure}

This study area comprises Zambia and Ghana (Figure~\ref{fig_study_area}). Ghana has two rainfall regimes (unimodal in the Savana zone, and bimodal in the Forest and Coastal zones) while Zambia has a unimodal rainfall regime.

\begin{figure}[H]
	\subfloat[]{
		\begin{minipage}[1\width]{
				0.4\textwidth}
			\centering
			\includegraphics[width=1\textwidth]{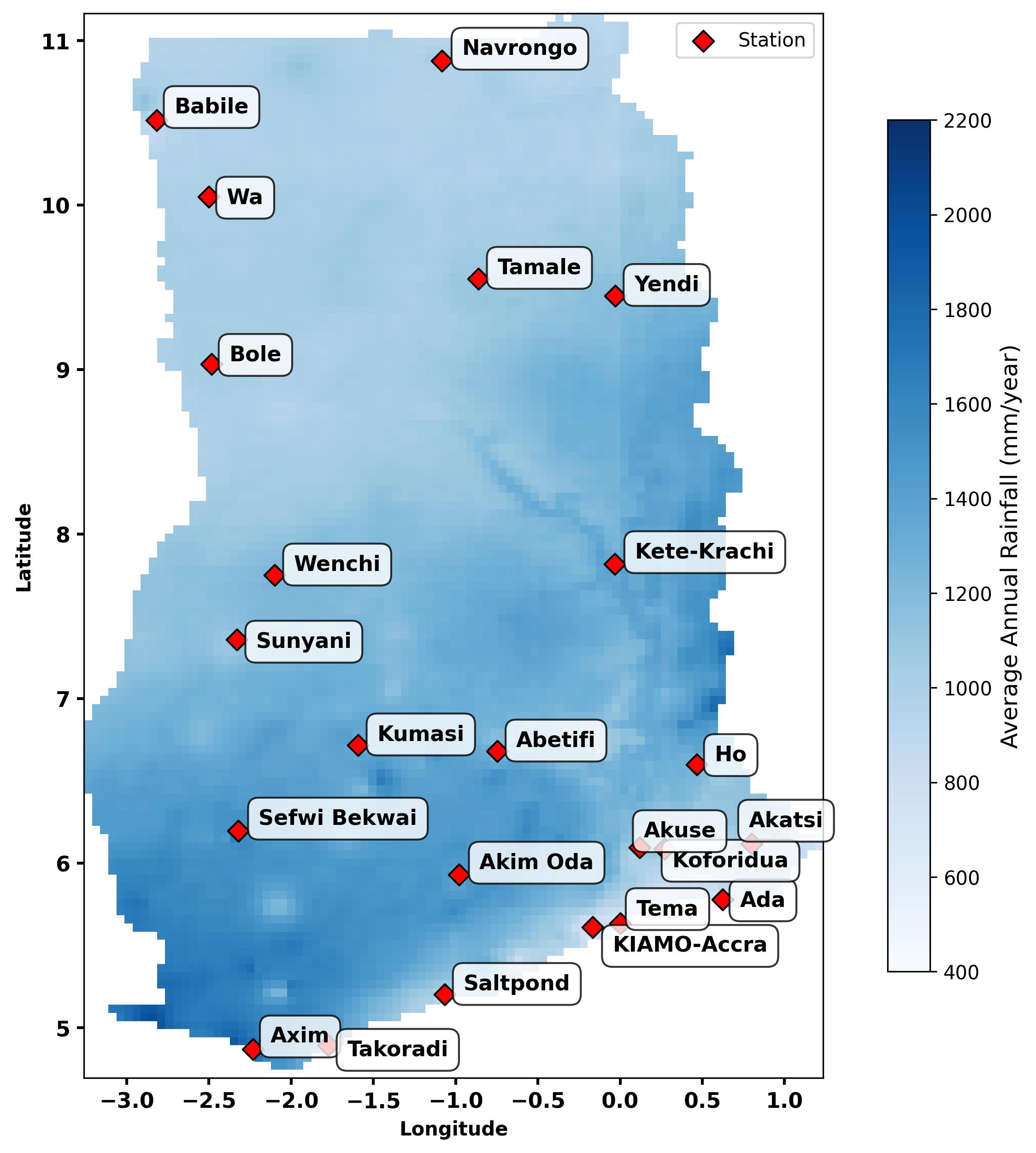}
	\end{minipage}}
	\hfill 	
	\subfloat[]{
		\begin{minipage}[1\width]{
				0.6\textwidth}
			\centering
			\includegraphics[width=1\textwidth]{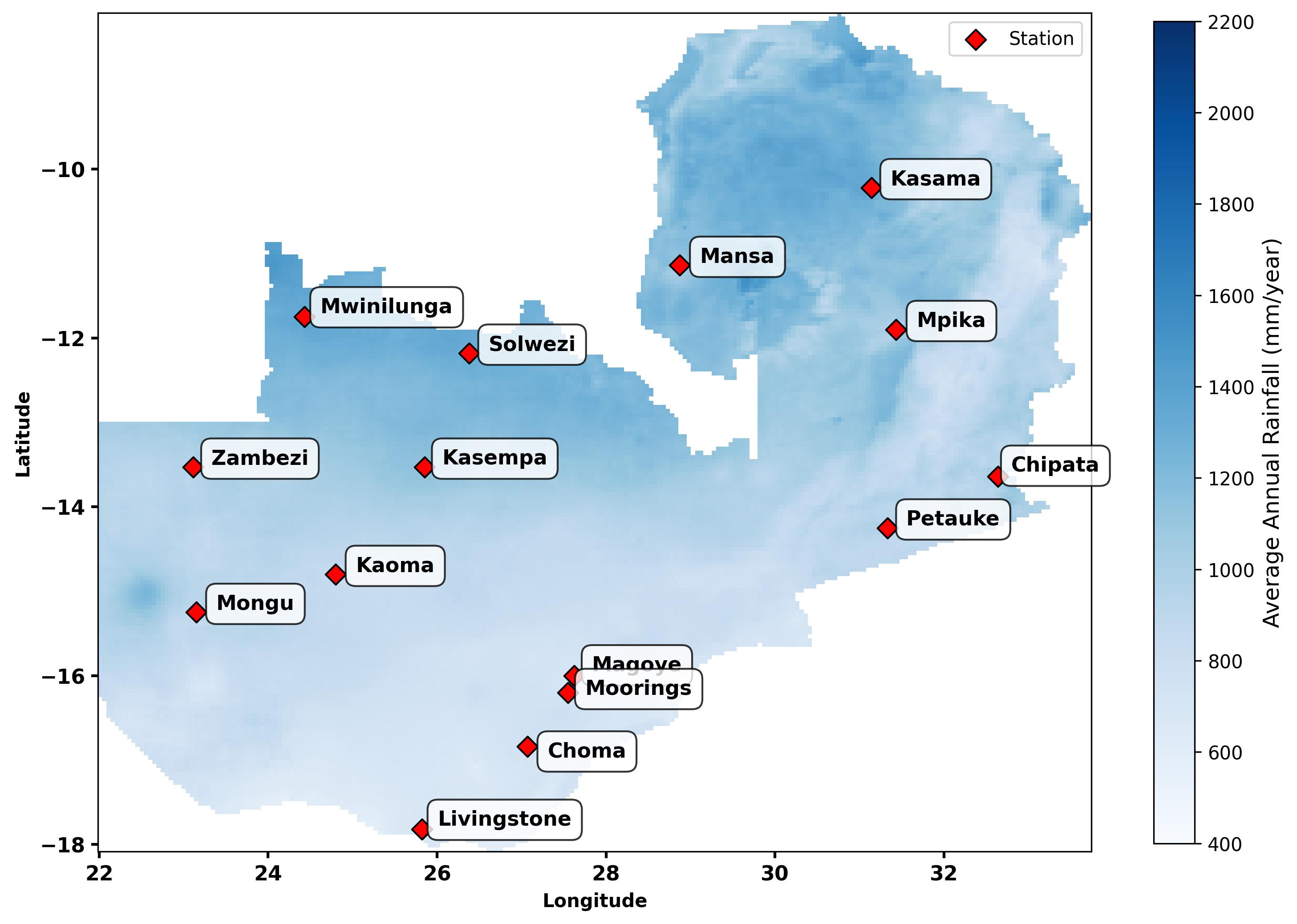}
	\end{minipage}}
	\caption{A spatial distribution map of mean annual rainfall over Ghana (\textbf{a}) and Zambia (\textbf{b}) for the period 1983--2022 using CHIRPS v2 data~\cite{Funk2014, Funk2015}.}\label{fig_av_rain}
\end{figure}

The spatial distribution of rainfall differs between Ghana and Zambia. While Ghana's central and southwestern coasts receive the highest precipitation, its northern and southeastern regions are relatively drier. Zambia exhibits the opposite pattern, with its northern parts receiving more rainfall than the south (Figure~\ref{fig_av_rain}). Refer to \cite{jain2007empirical, Kaczan2013, Maidment2017, hachigonta2008analysis, Oduro2024, boateng2021rainfall, Amekudzi2015-qk, torgbor2018rainfall, Bessah2022, Atiah2020-rz, Chisanga2023} for more details on climatology of these countries.

We used rain gauge data from 38 stations and processed pixel-to-point extracted SRE data. These stations are distributed across both countries (23 in Ghana and 15 in Zambia), covering all the climatic and agroecological zones in Ghana and Zambia, respectively (see Figure~\ref{fig_study_area}). 

A common difficulty when using station data is the possibility of errors or inconsistencies. To resolve this, every observational record was subjected to thorough quality assurance prior to its use in the research. These checks, as previously quality-controlled in~\cite{bagiliko2025}, involved identifying sequences of consecutive rainy days, repeated identical precipitation measurements, anomalously high values, months with no rainfall, and gaps in the data. Given the fundamental importance of data integrity, this effort demanded considerable time and close cooperation with the original data suppliers. Any measurements that did not pass the quality checks were treated missing. The study only incorporated stations that retained a minimum of 70\% of their data after this process. Furthermore, when calculating annual number of rainy days, any year containing fewer than 355 days of valid observations was omitted from analysis.

The seven SREs were: TAMSAT v3.1 (hereafter TAMSAT;~\cite{Maidment2017, Tarnavsky2014}), ERA5 (the fifth-generation ECMWF reanalysis~\cite{Hersbach2020, Bell2021}), CHIRP and its station-blended counterpart CHIRPS v2, referred here as CHIRPS~\cite{Funk2014, Funk2015}, the ECMWF Agrometeorological reanalysis (AgERA5;~\mbox{\cite{boogaard2020agrometeorological}}), PCDR~\cite{Ashouri2015}, and ENACTS~\cite{Dinku2017, Dinku2022}. These SREs uniformly possess long historical records (>30 years), high spatial resolution (a minimum of $0.25^{\circ}$), and fine temporal resolution (daily or sub-daily), while also encompassing Africa or portions of the study region (see Table~\ref{tab_satdata}). 

\begin{table}[h]
	\centering
	\caption{Details of SREs used in the study}
	\label{tab_satdata}
	\small
	\begin{tabular}{@{}lcccc@{}}
		\toprule
		\textbf{Product} & \textbf{Coverage} & \textbf{Period} & \textbf{Resolution} & \textbf{Reference} \\
		\midrule
		CHIRPS         & Global          & \makecell[c]{1981--\\present} & 0.05° daily        & \cite{Funk2014, Funk2015} \\
		CHIRP          & Global          & \makecell[c]{1981--\\present} & 0.05° daily        & \cite{Funk2014, Funk2015} \\
		TAMSAT         & Africa          & \makecell[c]{1983--\\present} & 0.0375° daily      & \makecell[c]{\cite{Maidment2017},\\ \cite{Tarnavsky2014},\\\cite{Hersbach2020}} \\
		ERA5           & Global          & \makecell[c]{1940--\\present} & 0.25° hourly       & \makecell[c]{\cite{Bell2021}\\\cite{Hersbach2020}} \\
		AGERA5         & Global          & \makecell[c]{1979--\\present} & 0.1° daily         & \cite{boogaard2020agrometeorological}\\
		ENACTS         & \makecell[c]{Selected\\countries} & \makecell[c]{1981--\\present} & 0.0375° daily    & \cite{Dinku2017, Dinku2022} \\
		PCDR           & Global          & \makecell[c]{1983--\\present} & 0.25° daily        & \cite{Ashouri2015} \\
		\bottomrule
	\end{tabular}
\end{table}

\subsection{Bias correction methods}
This study evaluates a comprehensive suite of BC methods: LOCI, QM, SVR, GPR. The
methods were chosen to cover mean-based (LOCI), distribution-based (QM), and ML-based
approaches (SVR and GPR).

\begin{figure}[H]
	\subfloat[]{
		\begin{minipage}[1\width]{
				0.5\textwidth}
			\centering
			\includegraphics[width=1\textwidth]{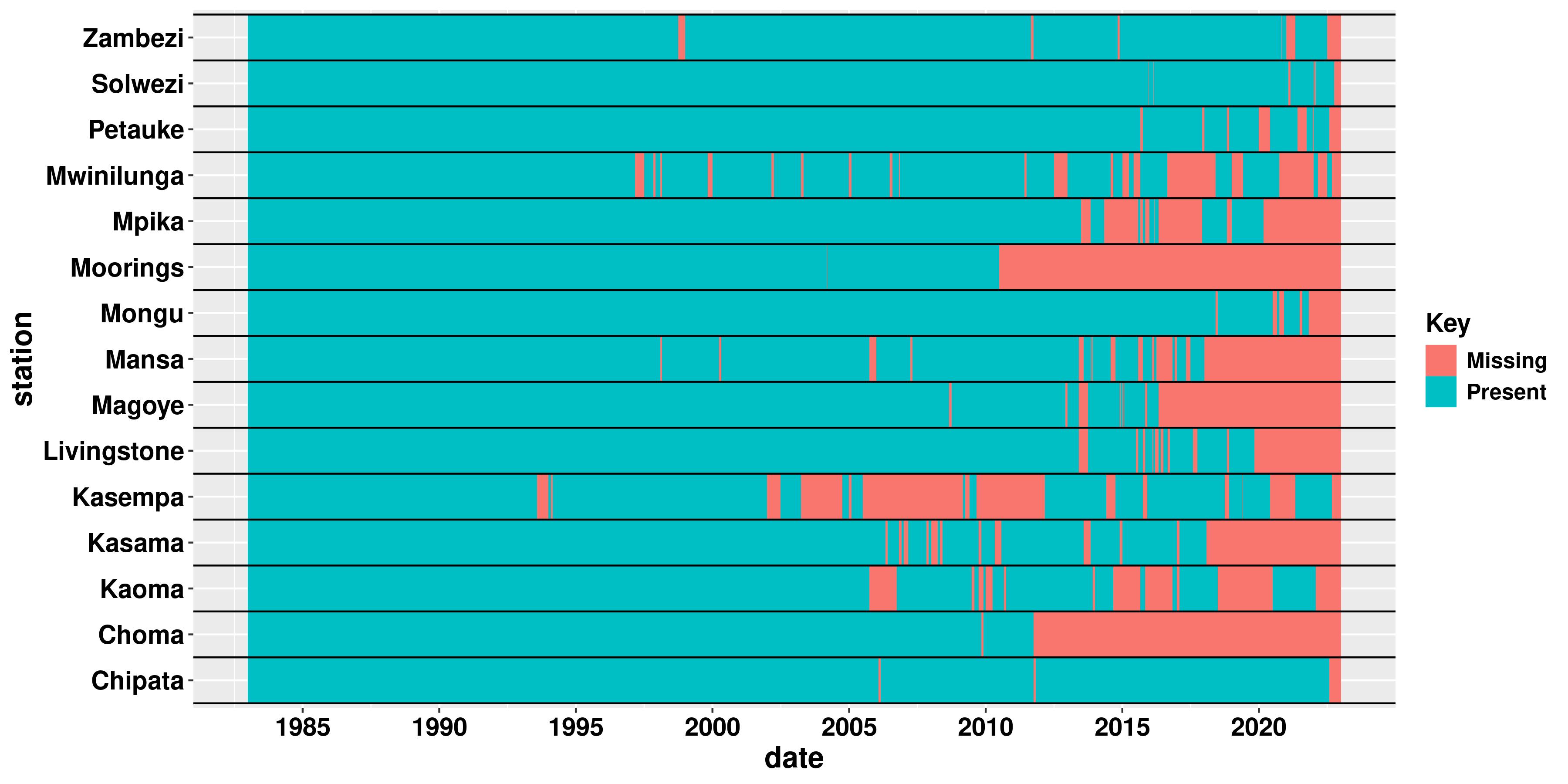}
	\end{minipage}}
	\hfill 	
	\subfloat[]{
		\begin{minipage}[1\width]{
				0.5\textwidth}
			\centering
			\includegraphics[width=1\textwidth]{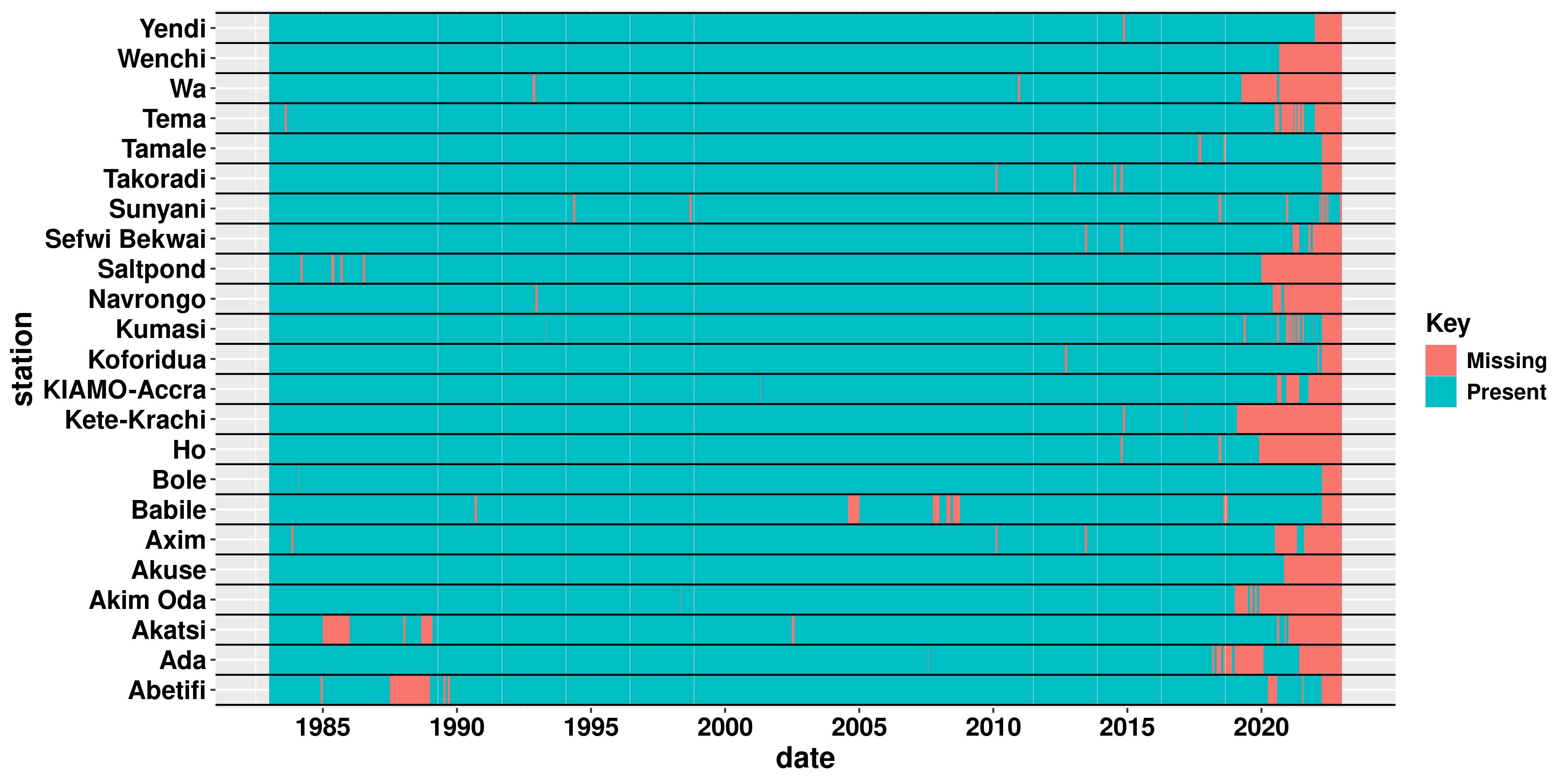}
	\end{minipage}}
	\caption{Inventory plot of the station data used in the study for both Zambia (\textbf{a}) and Ghana (\textbf{b}). \label{fig_qc}}
\end{figure}

Our analysis used a robust data splitting strategy spanning the historical period from 1983 to 2022 (depending on the length of observed records, as detailed in Figure \ref{fig_qc}). To ensure a stable and representative distribution of rainfall characteristics between the fitting (training) and testing datasets, we employed an odd-even year split: odd years were used for training (or fitting) while even years were used for testing.

\begin{figure}[H]
	\centering
	\includegraphics[width=.6\linewidth]{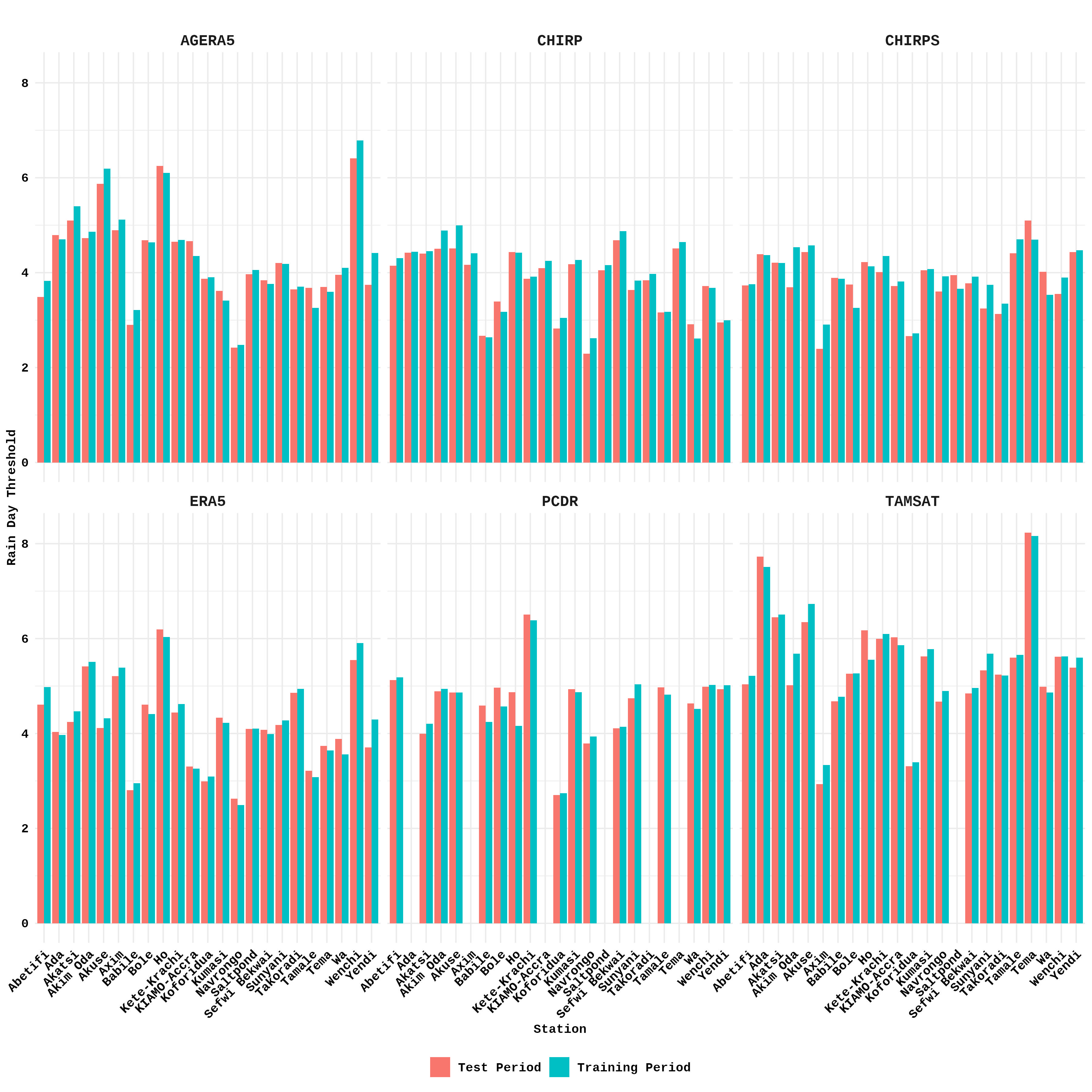}
	\caption{A grouped bar chart showing the fitted rain day thresholds of the SREs for the training period (odd years within 1983--2022) applied on the test period vs the actual rain day threshold for the test period (even years in the 1983--2022 period) across the stations and SREs in Ghana. \label{fig_temp_split2}} 
\end{figure}

\begin{figure}[H]
	\centering
	\includegraphics[width=.6\linewidth]{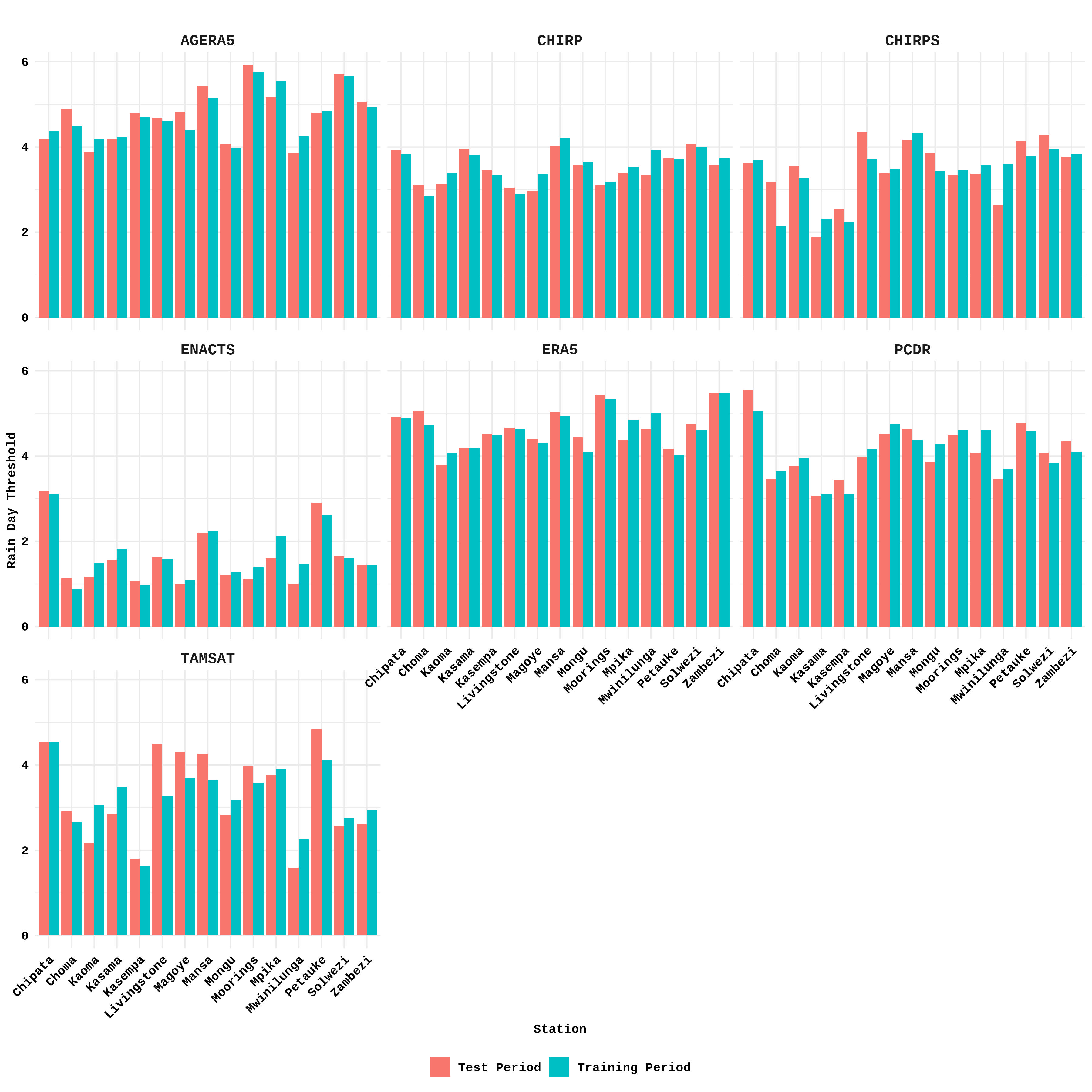}
	\caption{A grouped bar chart showing the fitted rain day thresholds of the SREs for the training period (odd years within 1983--2022) applied on the test period vs the actual rain day threshold for the test period (even years in the 1983--2022 period) across the stations and SREs in Zambia. \label{fig_temp_split2_zm}} 
\end{figure}

This approach was adopted specifically because it resulted in a greater stability of the rain day threshold between the training and testing subsets (Figures
~\ref{fig_temp_split2} and \ref{fig_temp_split2_zm}).

By using the odd-even year split, we reduced the effects of non-stationarity over time (Figures~\ref{fig_temp_split2} and \ref{fig_temp_split2_zm}). This approach ensures that both the training and testing datasets contain observations from across the entire 40-year period, effectively mixing data from earlier, middle, and later parts of the record.

\begin{figure}[H]
	\centering
	\includegraphics[width=.6\linewidth]{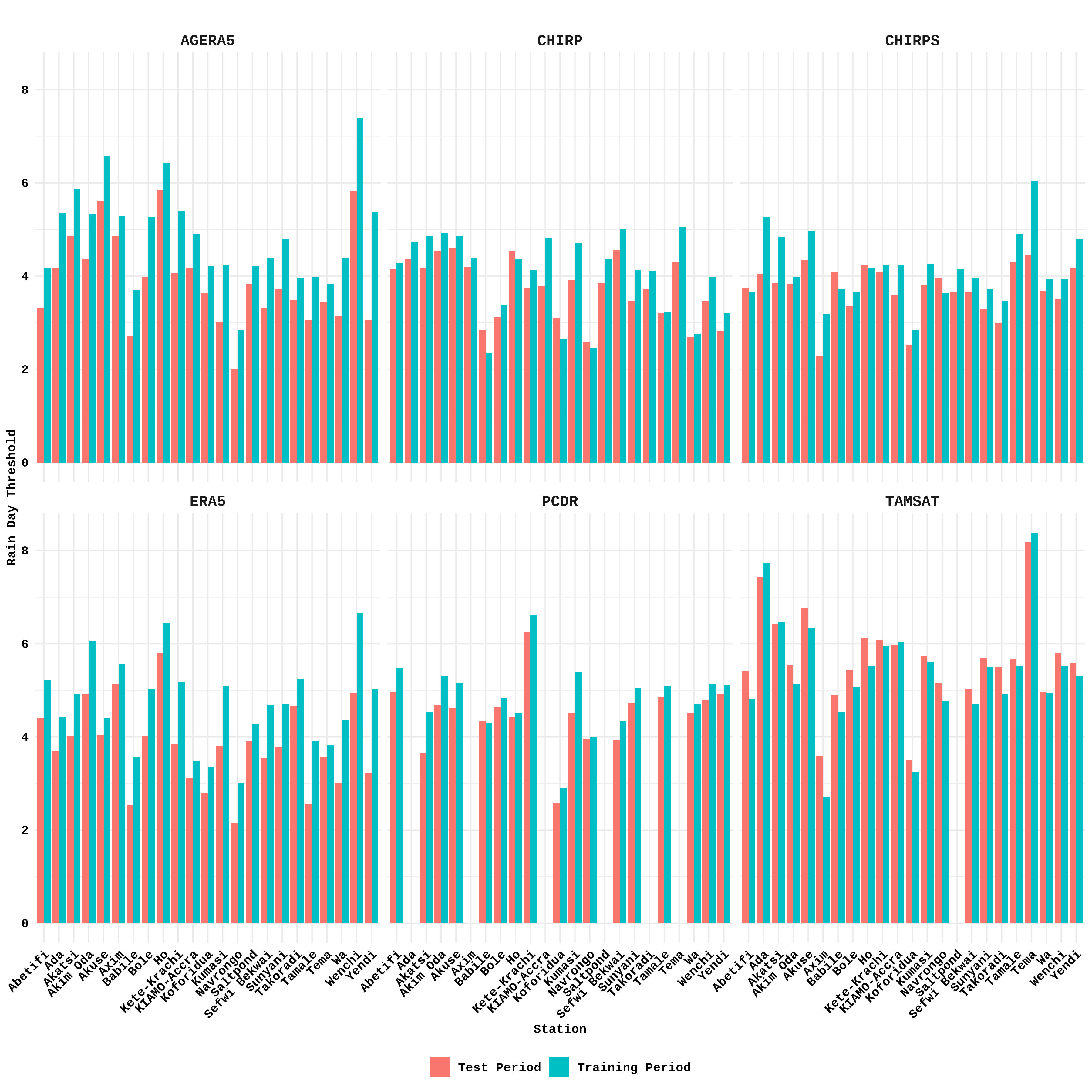}
	\caption{A grouped bar chart showing the fitted rain day thresholds of the SREs for the training period (1983--2000) applied on the test period vs the actual rain day threshold for the test period ($\geq$
		2001) across the stations in Ghana.  \label{fig_temp_split1}} 
\end{figure}

\begin{figure}[H]
	\centering
	\includegraphics[width=.6\linewidth]{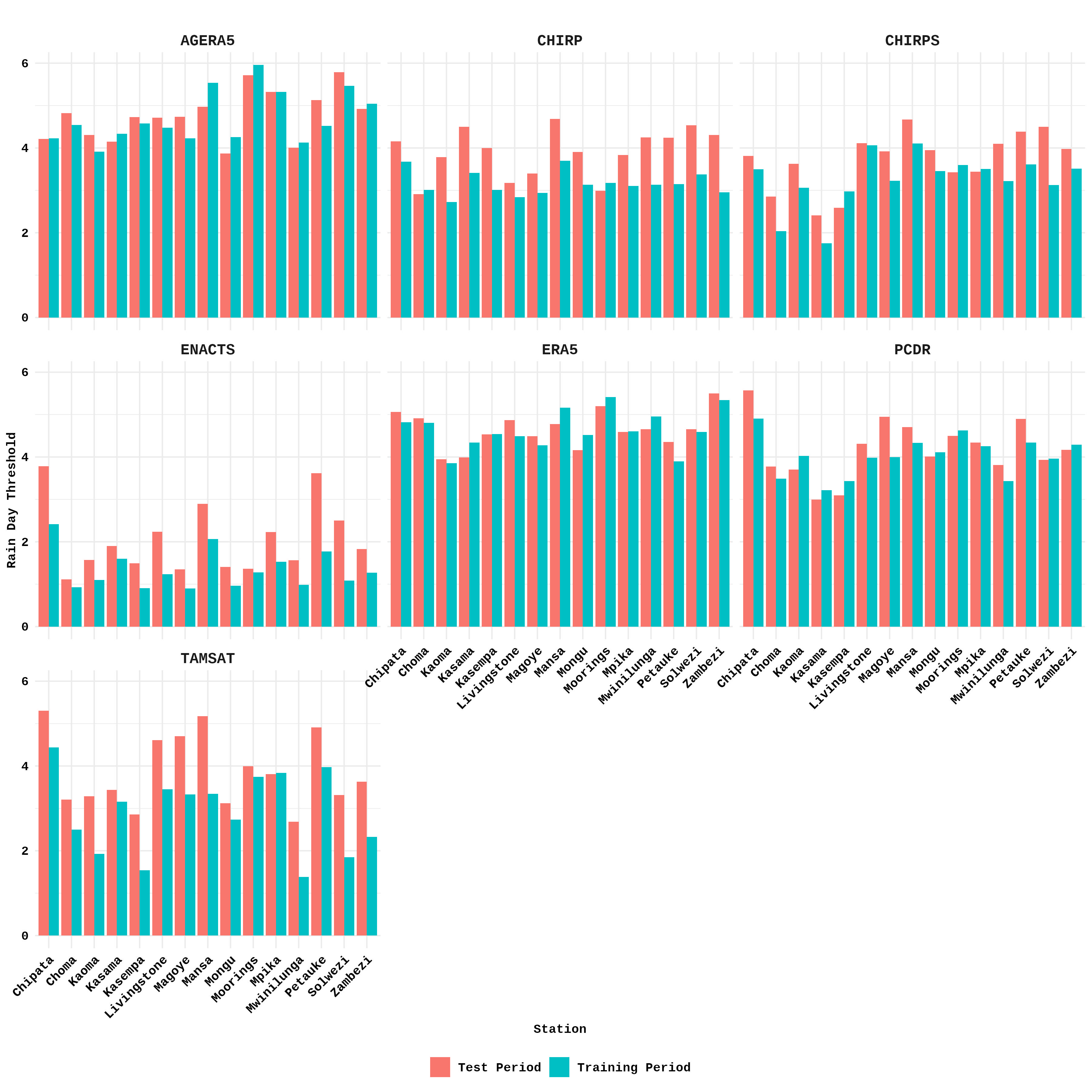}
	\caption{A grouped bar chart showing the fitted rain day thresholds of the SREs for the training period (1983--2000) applied on the test period vs the actual rain day threshold for the test period ($\geq$
		2001) across the stations in Zambia.  \label{fig_temp_split1_zm}}
\end{figure}

An initial attempt to use a continuous temporal split, where the period \mbox{1983--2000} served as training and 2001--2022 as testing (an approach used in some previous works~\mbox{\cite{Li2023, DHAWAN2024e40352}}), revealed a high discrepancy in key rainfall characteristics (Figures~Figure~\ref{fig_temp_split1} and \ref{fig_temp_split1_zm}), especially for Ghana (Figure~\ref{fig_temp_split1}), particularly the rain day thresholds, between the two periods. This discrepancy is likely attributable to the significant influence of climate change and other long-term factors on rainfall patterns.

An alternative strategy involves training models at one station and testing their performance on a neighbouring station (a spatial transferability approach). Given the significant distances and sparse location of stations in many African countries, including our study area, this spatial transfer approach was deemed unsuitable for our context.
\vspace{-6pt}
\vspace{-12pt}

\vspace{-12pt}

\vspace{-12pt}

%
%

For the ML methods, we trained the models using additional predictors, as detailed in Table~\ref{tab_vars}. These variables were selected due to their ready availability via the Google Earth Engine API and their comprehensive coverage of our study areas.

\begin{table}[H]
	\caption{Addition predictors used for machine learning, obtained from \cite{Copernicus2020AgERA5}}
	\label{tab_vars}
	\begin{tabularx}{\columnwidth}{XX}
		\toprule
		Variable & Description \\
		\midrule
		Wind\_Speed\_10m\_Mean\_24h & Wind speed \\
		Dew\_Point\_Temperature\_2m\_Mean\_24h & Dewpoint temperature, 2m \\
		Temperature\_Air\_2m\_Max\_24h & Maximum temperature, 2m \\
		Temperature\_Air\_2m\_Mean\_24h & Mean temperature, 2m \\
		Temperature\_Air\_2m\_Min\_24h & Minimum temperature, 2m \\
		Temperature\_Air\_2m\_Max\_Day\_Time & Max temperature, 2m, daytime \\
		Temperature\_Air\_2m\_Mean\_Day\_Time & Mean temperature, 2m, daytime \\
		Temperature\_Air\_2m\_Mean\_Night\_Time & Mean temperature, 2m, nighttime \\
		Cloud\_Cover\_Mean\_24h & Cloud cover \\
		ReferenceET\_PenmanMonteith\_FAO56 & FAO56 Reference ET \\
		Solar\_Radiation\_Flux & Downward solar radiation \\
		Vapour\_Pressure\_Mean\_24h & Vapour pressure \\
		\bottomrule
	\end{tabularx}
\end{table}

The BC methods are presented as follows:

\subsubsection{LOCI}\label{loci_sec}
LOCI~\cite{Schmidli2006} first calculates a scale factor, defined in 
(\ref{loci_eq}).
\begin{equation}
	s_{m} = \frac{\text{mean}(x_{i} |x_{i} \geq T^x_m) - T^x_m}{\text{mean}(y_{i} |y_{i} \geq T^y_m) - T^y_m},
	\label{loci_eq}
\end{equation}
\noindent
where $x_i$ is the station rainfall value on day $i$ in month $m$, $y_i$ is SRE rainfall value on day $i$ in month $m$, $T^y_m$ is the rain day threshold of the SRE in month $m$, with rain day threshold $T^x_m = 0.85$~\cite{STERN2011}.  

The threshold $T^y_m$ was obtained through quantile matching:
\begin{equation}
	T^y_m = F^{-1}_{y_m}\left(F_{x_m}(T^x_m)\right)
	\label{eq:threshold}
\end{equation}
\noindent
where $F_{x_m}$ is the Empirical cumulative distribution function (CDF) of station rainfall in month $m$, $F_{y_m}$ is the empirical CDF of SRE rainfall in month $m$, and $F^{-1}_{y_m}$ is the inverse CDF of SRE rainfall.

This ensures equivalent wet-day probabilities:
\begin{equation}
	P(y_i \geq T^y_m) = P(x_i \geq T^x_m)
\end{equation}

Finally, the scale factor is used to obtain $y'_i$, the bias-corrected SRE rainfall value on day $i$ given in (\ref{loc_bin}).
\begin{equation}
	y'_i= 
	\begin{cases}
		0,& \text{if } y_i\leq T^y_m\\
		T^x_m + s_{m}(y_i - T^y_m),              & \text{otherwise}
	\end{cases}
	\label{loc_bin}
\end{equation}

We note that the corrected values $y'_i$ can become unrealistically large if $s_m > 1$ and $y_i$ significantly exceeds $T^y_m$. However, we had various instances of $s > 1$ at many of the stations (Tables~S2 and S3), prohibiting the direct application of the original LOCI formulation of $y'_i$ at these locations.

To address this, we propose a constrained version of $y'_i$ as follows:

\begin{equation}
	y'_i = 
	\begin{cases}
		0, & \text{if } y_i \leq T^y_m \\
		\min\left(T^x_m + s_{m}(y_i - T^y_m), \text{Rmax}_m\right), & \text{otherwise}
	\end{cases}
	\label{loc_const}
\end{equation}
where $\text{Rmax}_m$ is some reasonable maximum rainfall value for the location (and time) of interest, based on historical data. In our case, the historical maximum observed station rainfall in month m (based on the training period) was used. The inclusion of $\text{Rmax}_m$ serves as a physical plausibility check. Standard LOCI assumes a constant linear relationship across all rainfall intensities; however, when $s_m > 1$, this relationship can lead to extreme overestimation during high-intensity events. By setting $\text{Rmax}_m$ to the historical maximum of the training period, we ensure that corrected values remain within the observed natural variability of the region, thereby avoiding the generation of artifactual extremes caused by the interaction of high SRE values and large scaling factors. We acknowledge that using a historical maximum as a cap is conservative and could theoretically suppress an unprecedented extreme event. However, the linear assumption of LOCI can break down at the high end of the distribution. We argue that capping at a documented historical maximum provides a more reliable estimate for hydrological modeling than allowing unconstrained linear inflation, which can introduce unrealistic flooding signals.

\subsubsection{QM}
The QM method aligns the probability distribution of the rainfall values from the SRE with the distribution of the observed data~\cite{Gudmundsson2012}. This is achieved by ensuring that the cumulative distribution function (CDF) of SRE data and observed data are consistent~\cite{Gudmundsson2012}. The rain day frequency was first adjusted with $T^y_m$ obtained in Section \ref{loci_sec}. Gamma cumulative distribution functions were then fitted to both the SRE data and corresponding station data. The bias-corrected daily rainfall estimates of the SRE data were obtained by using the inverse CDF of the observed data as shown in (\ref{qm_eq}). 
\begin{equation}
	P_{bc} = F_{o}^{-1}(F_{SRE}(P_{SRE})),
	\label{qm_eq}
\end{equation}
where $P_{bc}$ is the bias-corrected daily rainfall values of the SRE, $F_{o}^{-1}()$ is the inverse CDF of $P_o$ (the observed data) known as the quantile function~\cite{10.2166/wcc.2020.261}, and $F_{SRE}()$ is the CDF of the SRE data ($P_{SRE}$).

\subsubsection{GPR}\label{sect_gpr}
A Gaussian Process (GP) is defined by a mean function \( m(x) \) and a covariance function \( k(x, x') \), for inputs \( x \) and \( x' \)~\cite{rasmussen2006gaussian, Schulz2018}. Given a training dataset 
\( \mathbf{X} \) consisting of input vectors $x_{i}$ and their corresponding target values $y_{i}$, and a test set \( \mathbf{X_*} \) for which we want to predict the corresponding outputs $f_{*}$, the joint distribution of the observed target values \( \mathbf{y} \) and the function values \( \mathbf{f}_* \) at the test points under the Gaussian Process (GP) prior is given by:
\begin{equation}
	\begin{pmatrix}
		\mathbf{y} \\
		\mathbf{f}_*
	\end{pmatrix}
	\sim \mathcal{N}\left(
	\mathbf{0},
	\begin{bmatrix}
		K(\mathbf{X}, \mathbf{X}) + \sigma_n^2 I & K(\mathbf{X}, \mathbf{X}_*) \\
		K(\mathbf{X}_*, \mathbf{X}) & K(\mathbf{X}_*, \mathbf{X}_*)
	\end{bmatrix}
	\right)
\end{equation}
where \( K(\mathbf{X}, \mathbf{X}) \) is the covariance matrix computed with the kernel function \( k \) for all pairs of training points, \( K(\mathbf{X}, \mathbf{X}_*) \) is the covariance matrix between the training points and the test points, \( K(\mathbf{X}_*, \mathbf{X}_*) \) is the covariance matrix for the test points, \( \sigma_n^2 \) represents the noise variance in the observations, and \( I \) is the identity matrix.

Given this joint distribution, the conditional distribution of the function values \( \mathbf{f}_* \) at the test points given the observed data \( \mathbf{y} \), the posterior distribution, is:
\begin{equation}
	\mathbf{f}_* | \mathbf{X}, \mathbf{y}, \mathbf{X}_* \sim \mathcal{N}(\boldsymbol{\mu}_*, \Sigma_*)
\end{equation}
where the posterior mean \( \boldsymbol{\mu}_* \) and covariance \( \Sigma_* \) are given by:

\begin{equation}
	\boldsymbol{\mu}_* = K(\mathbf{X}_*, \mathbf{X})[K(\mathbf{X}, \mathbf{X}) + \sigma_n^2 I]^{-1}\mathbf{y}
\end{equation}

\begin{equation}
	\Sigma_* = K(\mathbf{X}_*, \mathbf{X}_*) - K(\mathbf{X}_*, \mathbf{X})[K(\mathbf{X}, \mathbf{X}) + \sigma_n^2 I]^{-1}K(\mathbf{X}, \mathbf{X}_*)
\end{equation}

In our context, \( \mathbf{X} \) contains a number of vectors: the daily SRE estimates at a given station, the SRE values of the previous day, as well as the addition predictors from Table~\ref{tab_vars} used during training; \( \mathbf{y} \) represents the daily rainfall observations at a corresponding station used during training; and \( \mathbf{X_*} \) contains the daily SRE estimates at same station, the SRE values of the previous day, as well as the additional predictors from Table~\ref{tab_vars} used during testing. We used the Mat\'{e}rn kernel given by:

\begin{equation}\label{eqn:mtn}
	k(x, x') = \sigma^2 \frac{2^{1-\nu}}{\Gamma(\nu)} \left(\frac{\sqrt{2\nu} d}{\rho}\right)^\nu K_\nu\left(\frac{\sqrt{2\nu} d}{\rho}\right)
\end{equation}
\noindent
where $k(x, x')$ is the covariance function, $\sigma^2$ is the variance parameter, $\nu$ is a parameter that controls the smoothness of the kernel, $\Gamma(\nu)$ is the gamma function, $\rho$ is the length scale parameter, $d = \|x - x'\|$, and $K_\nu$ is the modified Bessel function of the second kind of order~$\nu$.


\textls[-15]{To account for varying error characteristics across different rainfall magnitudes, distinct GPR models were developed for specific rainfall intensity categories \mbox{(i.e., [0.85, 5) mm/day,} [5, 20) mm/day, [20, 40) mm/day, and $>$40 mm/day). These thresholds based were on~\cite{ZambranoBigiarini2017}. Prior to model training, all predictor variables were normalized using Min--Max Scaling applied separately within each intensity category for every SRE at each station. Crucially, the scaling parameters were determined exclusively from the training data and subsequently applied to the test set to prevent data leakage. Given the computationally intensive nature of the process, involving a maximum of $\text{\#stations} \times \text{\#SREs} \times \text{\#categories}$ models, an exhaustive grid search for optimal hyperparameters was infeasible. Consequently, the hyperparameters were fixed after empirical trials led to a suitable balance between fit and generalization. The final parameters adopted were $\nu = 3/2$ (corresponding to the Mat\'{e}rn kernel), $\sigma = 1.0$, $\rho = 2.0$, and $\sigma_n^2 = 0.10$. During training, if there were a small number of training samples (<10) in a category for an SRE at a station (as was the case for violent rains for many of the SREs), a model was not trained in which case the original SRE estimates were retained as the corrected data for that category. Negative predictions were converted to 0s. Due data scarcity within some of the intensity categories we did not perform any form of validation inside the training~period.}

\subsubsection{SVR}

\textls[-15]{SVR is a supervised ML technique based on Support Vector Machines initially developed by Vapnik and Chervonenkis in 1963~\cite{Cortes1995, GarcaFloriano2018}. The prediction function for SVR is given~by:}
\begin{equation}
	f(x) = \sum_{i=1}^{n} (\alpha_i - \alpha_i^*) k(x_i, x) + b
\end{equation}
where $\alpha_i$ and $\alpha_i^*$ are Lagrange multipliers obtained from the optimization process~\cite{10.1162/089976600300015565}, $k(x_i, x)$ is the kernel function evaluating the similarity between a new data point $x$ and the support vectors $x_i$, $n$ is the number of support vectors, and $b$ is the bias term. The radial basis function $k(x_i, x)$ defined in (\ref{eqn:rbf}) was used.
\begin{equation}\label{eqn:rbf}
	k(x, x') = v^2\exp\left(-\frac{\|x - x'\|^2}{2\lambda^2} \right)
\end{equation}
where $x, x'$ are vectors in the input space, $v^2$ is the kernel variance, $\|x - x'\|$ is the Euclidean distance between the vectors $x$ and $x'$, and $\lambda$ is the length scale parameter, which determines the smoothness of the function. 

Training the SVR means:
\begin{equation}
	\min_{w, b, \xi, \xi^*} \left( \frac{1}{2} \|w\|^2 + C \sum_{i=1}^n (\xi_i + \xi_i^*) \right)
\end{equation}
subject to the following constraints:
\begin{align}
	y_i - (w^T \phi(x_i) + b) &\leq \epsilon + \xi_i, \\
	(w^T \phi(x_i) + b) - y_i &\leq \epsilon + \xi_i^*, \\
	\xi_i, \xi_i^* &\geq 0 \quad \forall i.
\end{align}
\noindent
where $w$ is the weight vector, $\xi$ and $\xi^*$ are the slack variables, $\phi(x_i)$ represents the feature mapping of $x_i$, and $C$ is the regularization parameter controlling the trade-off between the flatness of the model and the amount up to which deviations larger than $\epsilon$ are tolerated.


As in the case of GPR, all predictor variables were normalized using Min--Max Scaling applied separately within each intensity category for every SRE at each station. For the SVR implementation using the RBF kernel, the following hyperparameters were selected after empirical tuning: $C = 100$; $\epsilon = 0.01$; and $\gamma = 0.5$). Negative predictions were converted to 0 s.

\subsection{Evaluation metrics} 
We evaluated the effectiveness of the BC methods on the SREs in capturing rainfall occurrence and intensity using independent test data. The assessment employed six complementary statistical metrics: Mean Error (ME), Pearson's correlation coefficient (Corr), Ratio of Standard Deviations (RSD), Root Mean Squared Error (RMSE), Mean Absolute Error (MAE), Nash-Sutcliffe efficiency (NSE) for both daily and annual scales. Probability of Detection (POD) was additionally used for evaluation at the daily temporal scale for assessing the ability of the SREs to detect dry days, heavy and violent rainfall events. These metrics collectively assess different dimensions of agreement between the corrected estimates, original SREs, and ground-based gauge observations. ME measures the average model bias, where positive and negative values signify systematic overestimation and underestimation of observed rainfall, respectively \cite{cli12110169}. The Corr quantifies the strength and direction of the linear relationship between the estimates and observations, ranging from $-1$ to $1$ \cite{Yang2016}. A value approaching $+1$ denotes a strong ability to capture temporal rainfall variability. RSD compares the variability of the estimates to the observations. An RSD value of 1 indicates perfect agreement, values >1 suggest over-dispersion, and values <1 indicate under-dispersion. The POD assesses the likelihood that a rainfall event identified by the gauge is also correctly detected by the SRE or its corrected version. The NSE is a normalized statistic that evaluates predictive skill. An NSE of 1 represents a perfect model, a value of 0 indicates predictions are as accurate as the mean of the observations, and values significantly below 0 suggest poor model performance. The RMSE is measure of error magnitude that is sensitive to large outliers due to the squaring of terms. A perfect score is 0. Finally, MAE provides a direct measure of average error magnitude and is less sensitive to extreme values than RMSE. Table \ref{tab:metrics} provides a summary of the metrics used in the evaluation.

\begin{table}[H]
	\centering
	\caption{Statistical metrics used for BC evaluation}
	\vspace{0.2cm}      
	\label{tab:metrics}
	\begin{tabular}{llll}
		\hline
		\textbf{Metric} & \textbf{Formula} & \textbf{Range} & \textbf{Optimal Value} \\ 
		\hline
		ME & $\frac{1}{n}\sum_{i=1}^n (P_i - O_i)$ & $(-\infty, +\infty)$ & 0 \\
		Corr & $\frac{\sum(P_i - \bar{P})(O_i - \bar{O})}{\sqrt{\sum(P_i - \bar{P})^2\sum(O_i - \bar{O})^2}}$ & [-1, 1] & 1 \\
		RSD & $\sigma_P/\sigma_O$ & $(0, +\infty)$ & 1 \\
		RMSE & $\sqrt{\frac{1}{n}\sum_{i=1}^n (P_i - O_i)^2}$ & $[0, +\infty)$ & 0 \\
		MAE & $\frac{1}{n}\sum_{i=1}^n |P_i - O_i|$ & $[0, +\infty)$ & 0 \\
		NSE & $1 - \frac{\sum(P_i - O_i)^2}{\sum(O_i - \bar{O})^2}$ & $(-\infty, 1]$ & 1 \\
		POD & $\frac{\text{Hits}}{\text{Hits} + \text{Misses}}$ & [0, 1] & 1 \\
		\hline
	\end{tabular}
	\vspace{0.4cm} 
	\footnotesize
	\begin{tabular}{@{}p{\textwidth}@{}}
		$P_i$: predicted value; $O_i$: observed value; $\bar{P},\bar{O}$: means of predicted and observed values respectively; $\sigma_P,\sigma_O$: standard deviations of predicted and observed values respectively; \\
		n: sample size; Hits: correct rainfall events detected; Misses: undetected rainfall events; False Alarms: rainfall event predicted but not observed.
	\end{tabular}
\end{table}

\section{Results}\label{sec3}

This section addresses the core research question, "how do BC methods applied to SREs contribute to the detection of rainfall events and the estimation of their intensities?" The results are presented at daily, seasonal, and annual scales to evaluate methodological performance across varying temporal aggregations.
\subsection{Performance of BC methods on daily scale}\label{daily_results}
\subsubsection{Overall performance of BC methods and SREs on daily rainfall amounts}\label{daily_overall}

\begin{figure}[H]
	\subfloat[]{
		\begin{minipage}[1\width]{
				0.5\textwidth}
			\centering
			\includegraphics[width=1\textwidth]{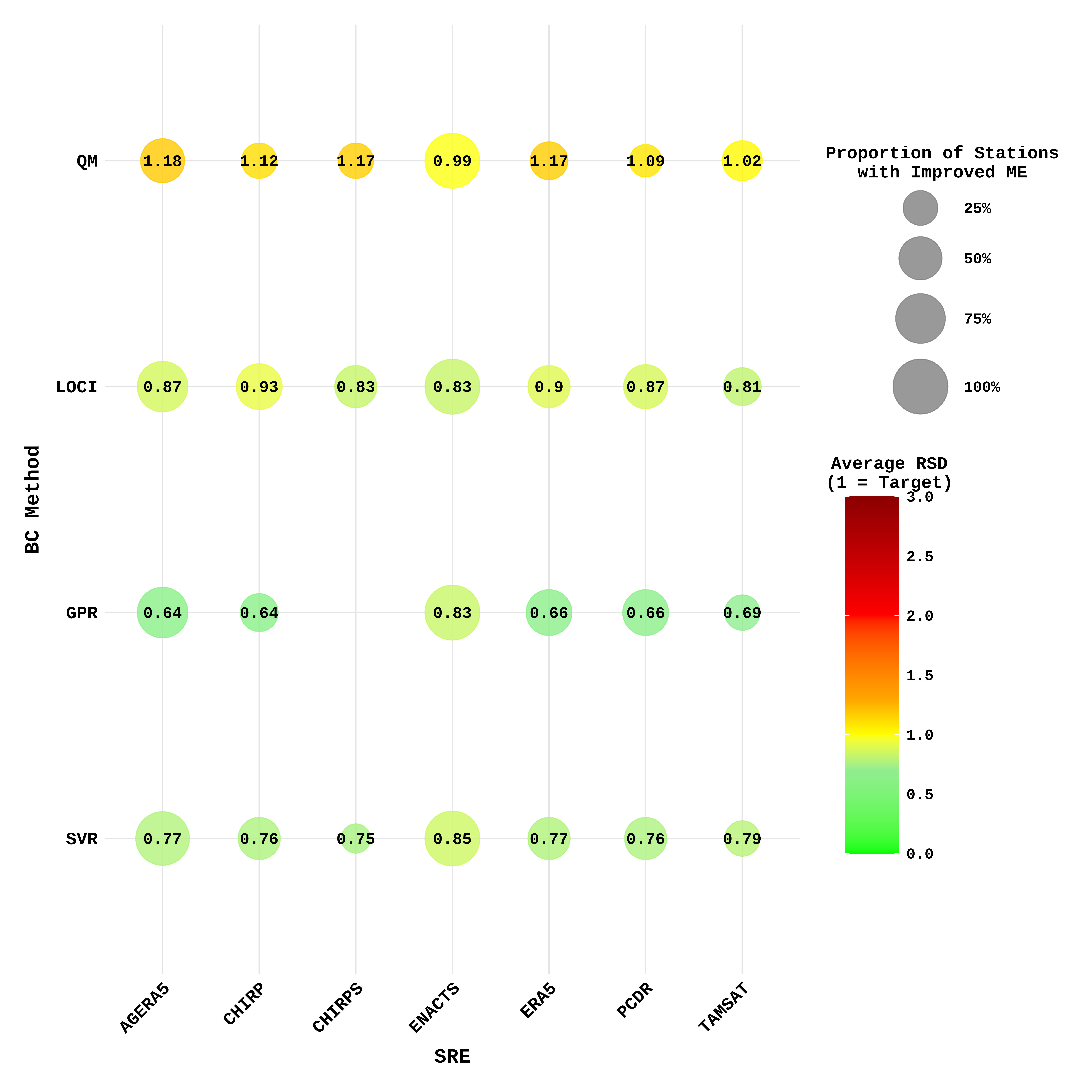}
	\end{minipage}}
	\hfill 	
	\subfloat[]{
		\begin{minipage}[1\width]{
				0.5\textwidth}
			\centering
			\includegraphics[width=1\textwidth]{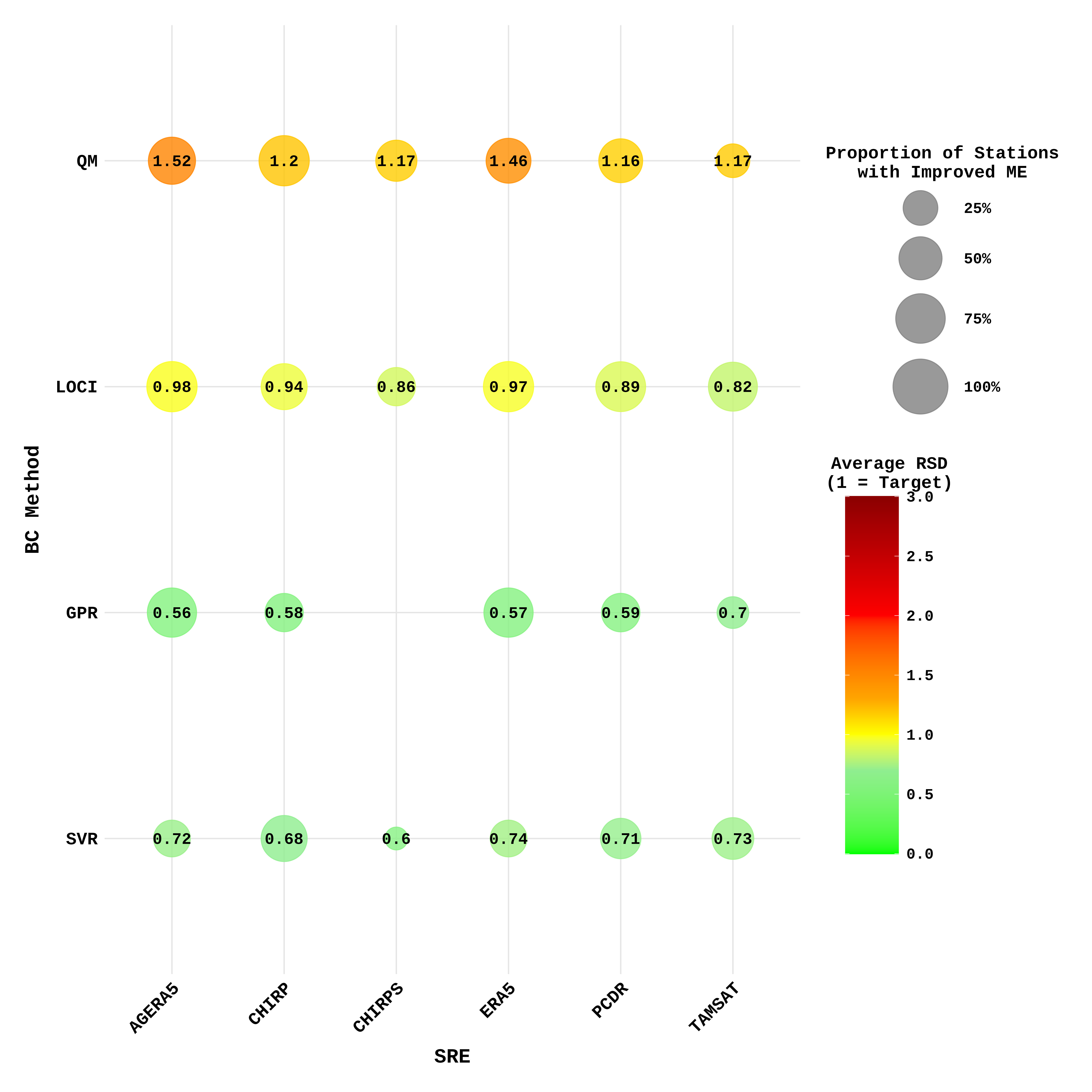}
	\end{minipage}}
	\caption{Bubble plots evaluating BC method performance in reducing ME on daily rainfall amounts for Zambia (a) and Ghana (b). Each bubble's position shows a BC method and SRE combination, its size shows the proportion of stations with reduced ME, and its colour shows the mean RSD at those stations. Missing bubbles indicate no improvement}\label{bc_me<sre_me_zm_gh}
\end{figure}

The efficacy of the BC methods was evaluated by their ability to reduce mean error (ME) in the SREs. Bubble plots in Figure~\ref{bc_me<sre_me_zm_gh}a,b display the proportion of stations in Zambia and Ghana, respectively, where a given BC method (y-axis) reduced the daily ME relative to the uncorrected SRE (x-axis). Bubble color represents the mean RSD (this mean is only for stations where there has been a reduction in the ME) across these stations, while bubble size corresponds to the proportion of stations. The mean RSD is also displayed numerically within each bubble. Missing bubbles indicate BC-SRE combinations that yielded no improvement.

Figure \ref{bc_me<sre_me_zm_gh}a revealed that the efficacy of the BC methods in Zambia varied substantially by SRE and station. When applied to ENACTS, almost all BC methods reduced the ME at a high proportion of stations (>70\%). Among them, QM and LOCI were particularly effective, successfully reducing ME for most SREs across many stations while maintaining reasonable variability (RSD close to 1), underscoring their superiority for capturing daily rainfall amounts. This was followed by SVR and then GPR. The latter two methods to perform better in Zambia than in Ghana in terms of terms reducing MEs at a larger proportion of stations while keeping RSD closer to 1.

\begin{figure}[H]
	\subfloat[]{
		\begin{minipage}[1\width]{
				0.5\textwidth}
			\centering
			\includegraphics[width=1\textwidth]{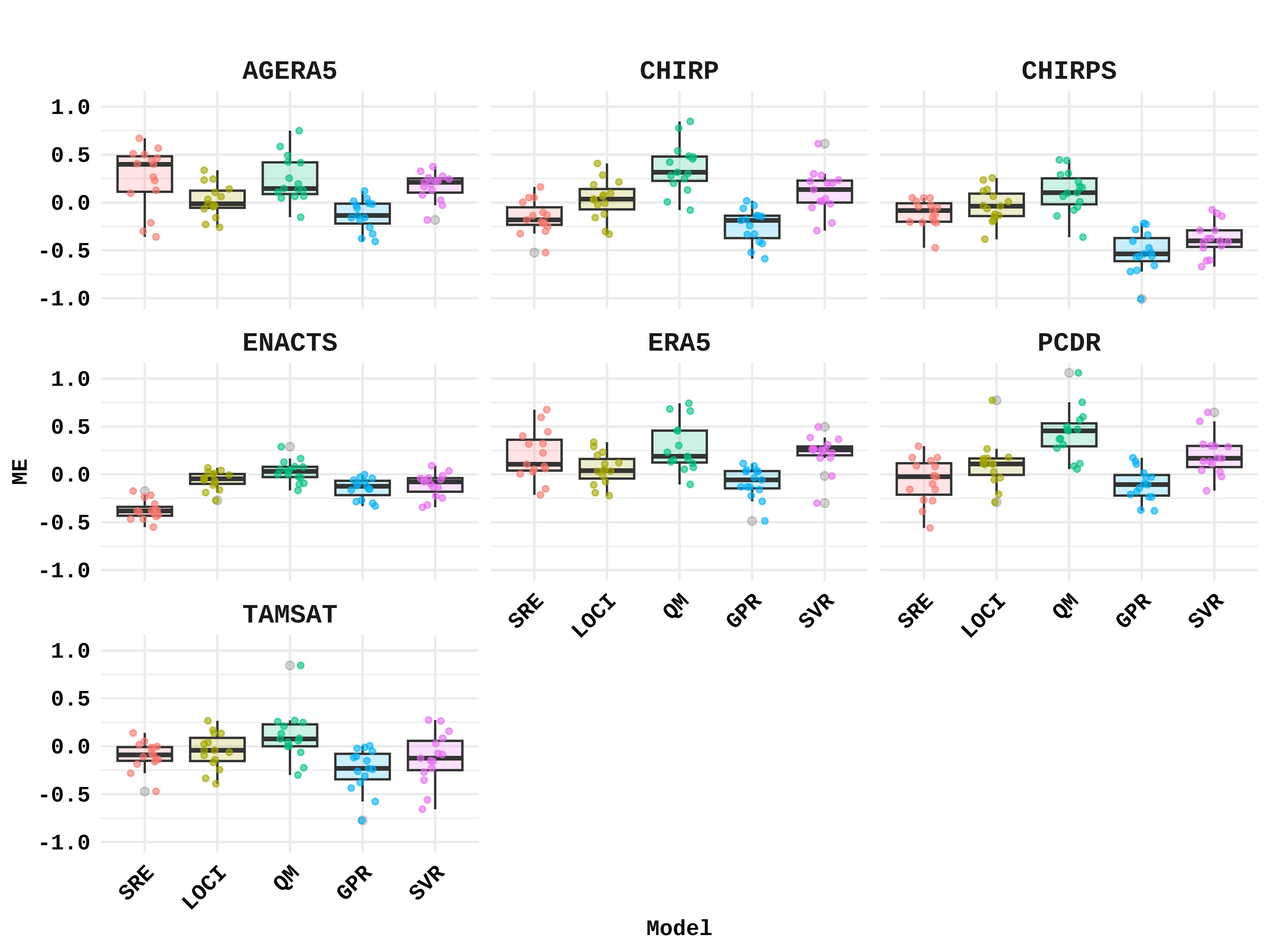}
	\end{minipage}}
	\hfill 	
	\subfloat[]{
		\begin{minipage}[1\width]{
				0.5\textwidth}
			\centering
			\includegraphics[width=1\textwidth]{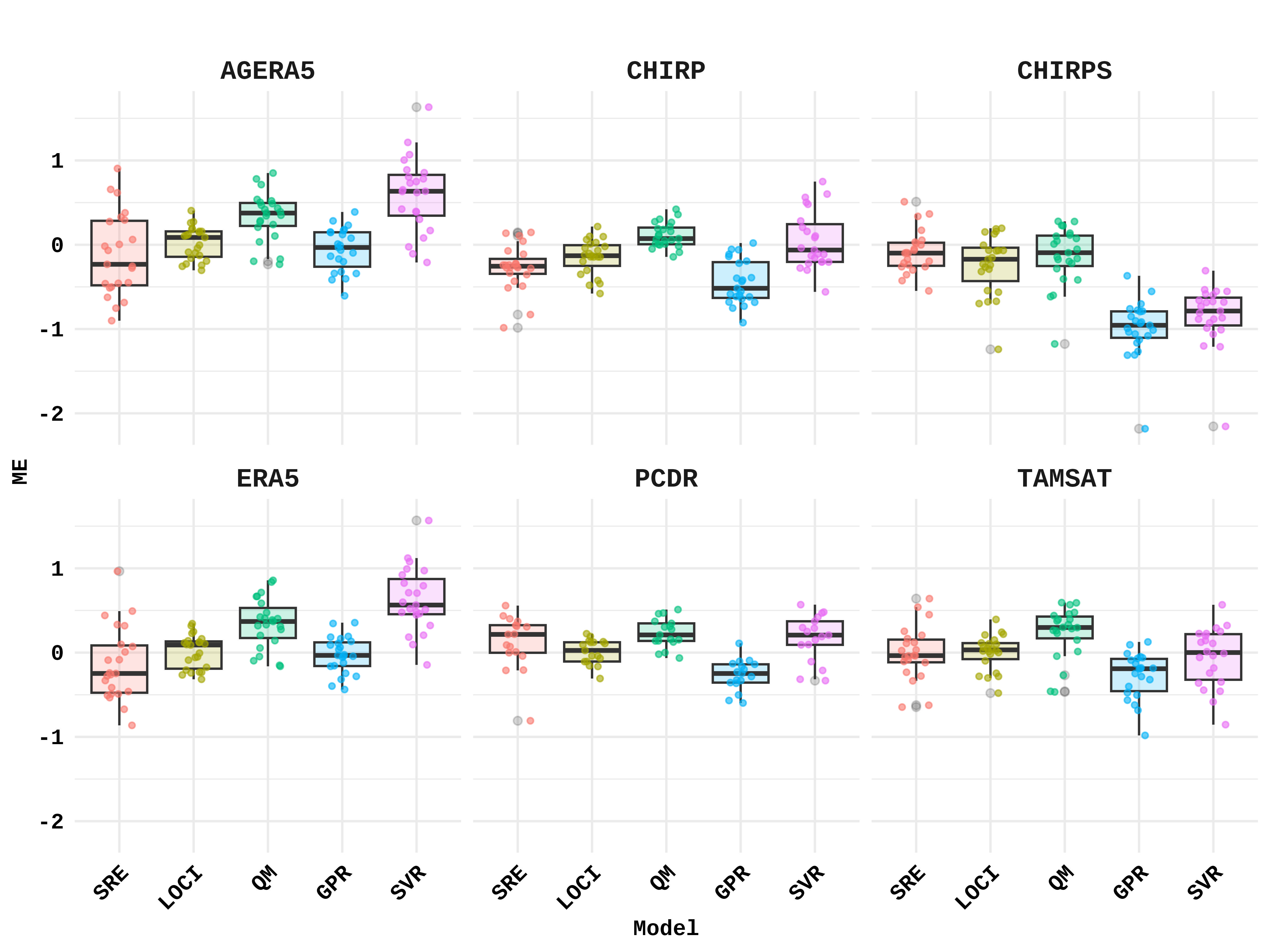}
	\end{minipage}}
	\caption{Mean errors (MEs) of the BC methods applied to SREs for (a) Zambia and (b) Ghana. Distributions are represented by box plots; individual data points are shown as jittered points}\label{av_me_pr_gh_zm}
\end{figure}

CHIRPS stood out as an SRE which most of the methods struggled to reduce the MEs, especially the ML methods (Figure \ref{bc_me<sre_me_zm_gh}). Particularly, GPR reduced MEs for some stations across all SREs except CHIRPS. Analysis of daily values indicated that GPR typically underestimated rainfall, leading to negative MEs and lower variability. This suggests GPR is most effective at stations where the uncorrected SREs had high positive biases. This explains its poor performance with CHIRPS and TAMSAT, which generally exhibited low positive or negative biases (Figure \ref{av_me_pr_gh_zm}(a)), as GPR's underestimation worsened their MEs.

The strong performance of BC methods on ENACTS is likely a characteristic of the ENACTS dataset itself, rather than the BC methods. Although uncorrected ENACTS had a negative ME, GPR and SVR still reduced this error, contrary to their behavior with other SREs, indicating that the initial distribution of ENACTS is more amenable to these corrections.

A similar pattern was observed in Ghana (Figure~\ref{bc_me<sre_me_zm_gh}b), where QM and LOCI again improved ME at many stations while preserving reasonable variability. However, for the reanalysis products (AGERA5 and ERA5), which had high positive uncorrected MEs (Figure~\ref{av_me_pr_gh_zm}b), QM significantly overestimated variability. GPR and SVR appeared to perform better in Zambia than in Ghana in terms reducing MEs at a larger proportion of stations while keeping RSD closer to 1.

The previous indicator, based solely on the reduction of ME, can be noisy, as a reported reduction or increase in ME might be negligible in magnitude. To reduce the effect of this noise, we introduced another indicator: whether the BC method brings the ME within an acceptable threshold, defined as less than 20\% of the observed mean daily rainfall. This indicator identifies substantial improvements together with minor, inconsequential changes (i.e. slight decrease or increase in ME). The selected threshold serves a specific quantitative purpose; however, its value is not fixed and may be adjusted based on the context or sensitivity requirements of a given application.

\begin{figure}[H]
	\subfloat[]{
		\begin{minipage}[1\width]{
				0.5\textwidth}
			\centering
			\includegraphics[width=1\textwidth]{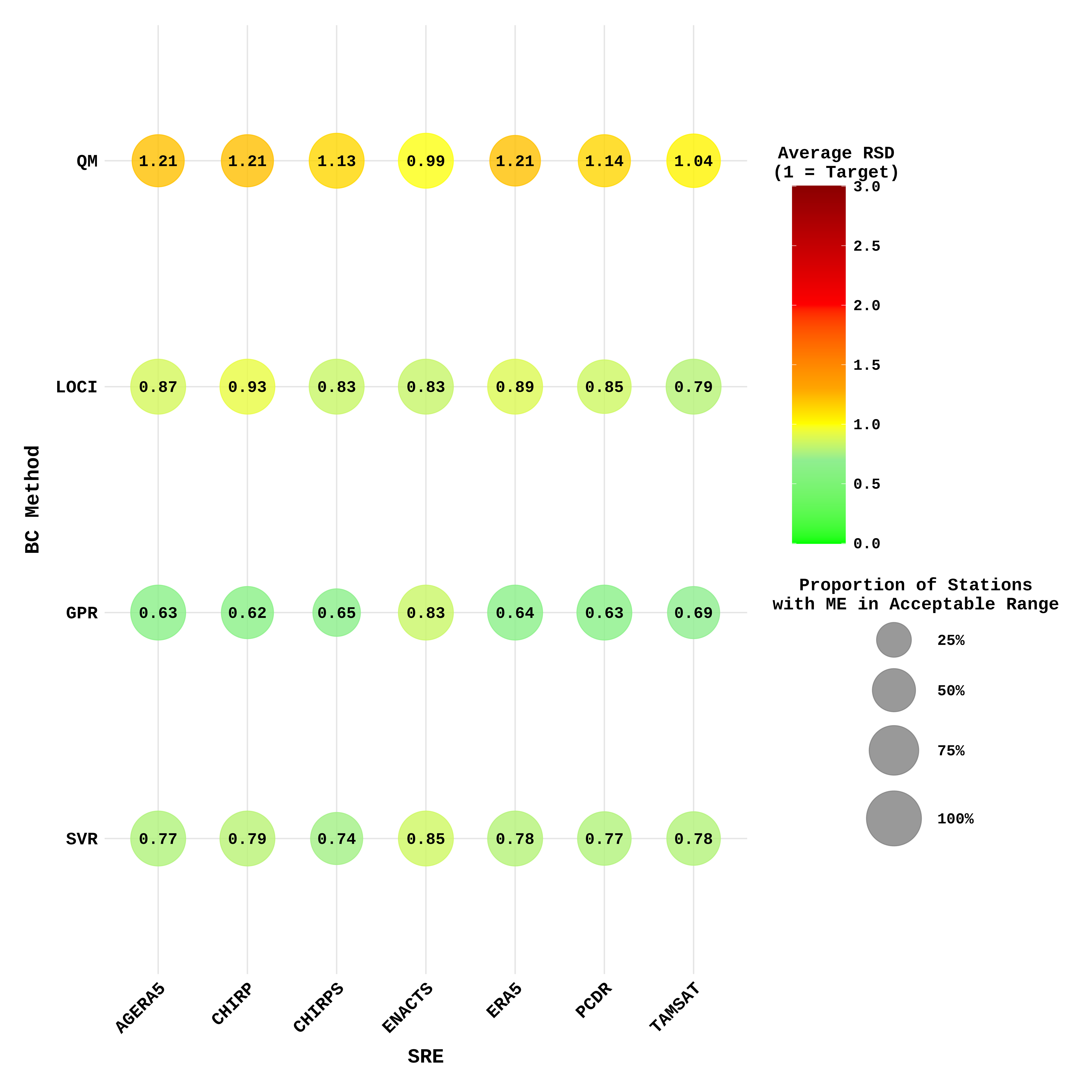}
	\end{minipage}}
	\hfill 	
	\subfloat[]{
		\begin{minipage}[1\width]{
				0.5\textwidth}
			\centering
			\includegraphics[width=1\textwidth]{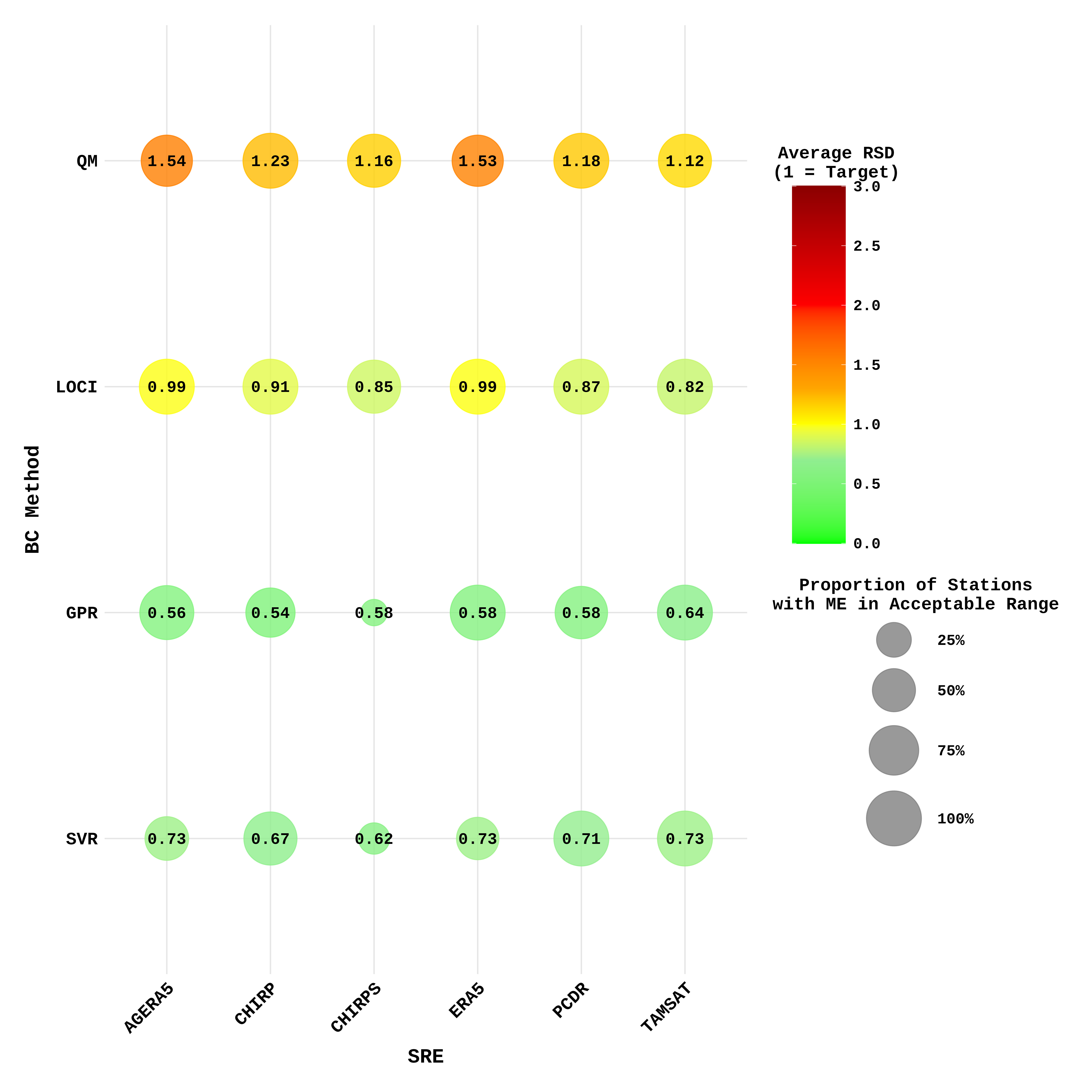}
	\end{minipage}}
	\caption{Proportion of stations in Zambia (a) and Ghana (b) where BC methods achieved an acceptable ME (< 20\% of observed mean rainfall). Bubble sizes correspond to the proportion of stations; colours indicate the mean RSD (also displayed inside the bubbles)}\label{bc_me_acceptable_zm_gh}
\end{figure}

Figures~\ref{bc_me_acceptable_zm_gh}(a) and (b) present bubble plots showing the proportion of stations in Zambia and Ghana, respectively, where the BC methods achieved an acceptable ME. In these plots, bubble sizes represent the proportion of stations, while bubble colours show the corresponding average RSD, with the mean RSD value displayed numerically inside each bubble.

In Zambia (Figure \ref{bc_me_acceptable_zm_gh}(a)), QM and LOCI consistently produced acceptable MEs at nearly all stations for all SREs, while also maintaining realistic variability (RSD close to 1), although QM overestimated variability for AGERA5 and ERA5. SVR followed suit by producing acceptable MEs at nearly all stations for all SREs but with lesser RSD values. GPR also achieved acceptable MEs for a almost all stations across SREs, but with lesser RSDs than SVR. This is consistent with its tendency to underestimate rainfall. ENACTS stood out as the SRE for which all BC methods achieved acceptable MEs at nearly all stations with reasonable variability. This performance appears to be an inherent property of the ENACTS dataset itself. 

In Ghana (Figure \ref{bc_me_acceptable_zm_gh}(b)), LOCI and QM produced acceptable MEs at a high proportion of stations for all SREs. However, QM tended to highly overestimate variability, an effect particularly pronounced for AGERA5 and ERA5 whose uncorrected versions mostly underestimated rainfall (negative MEs, see Figure~\ref{av_me_pr_gh_zm}), had low correlations and low variability (Table S1). The overestimation in variability by QM on these SREs is likely due to the introduction of high rainfall values on wrong days, leading to high positive biases and reduced correlations (Table S1). GPR and SVR also produced acceptable MEs at large proportion of the stations across SREs (except CHIRPS) albeit with low RSD values.  

\subsubsection{Detection of dry days}\label{dry_day_detection}
\begin{figure}[H]
	\centering
	\includegraphics[width=10.5 cm]{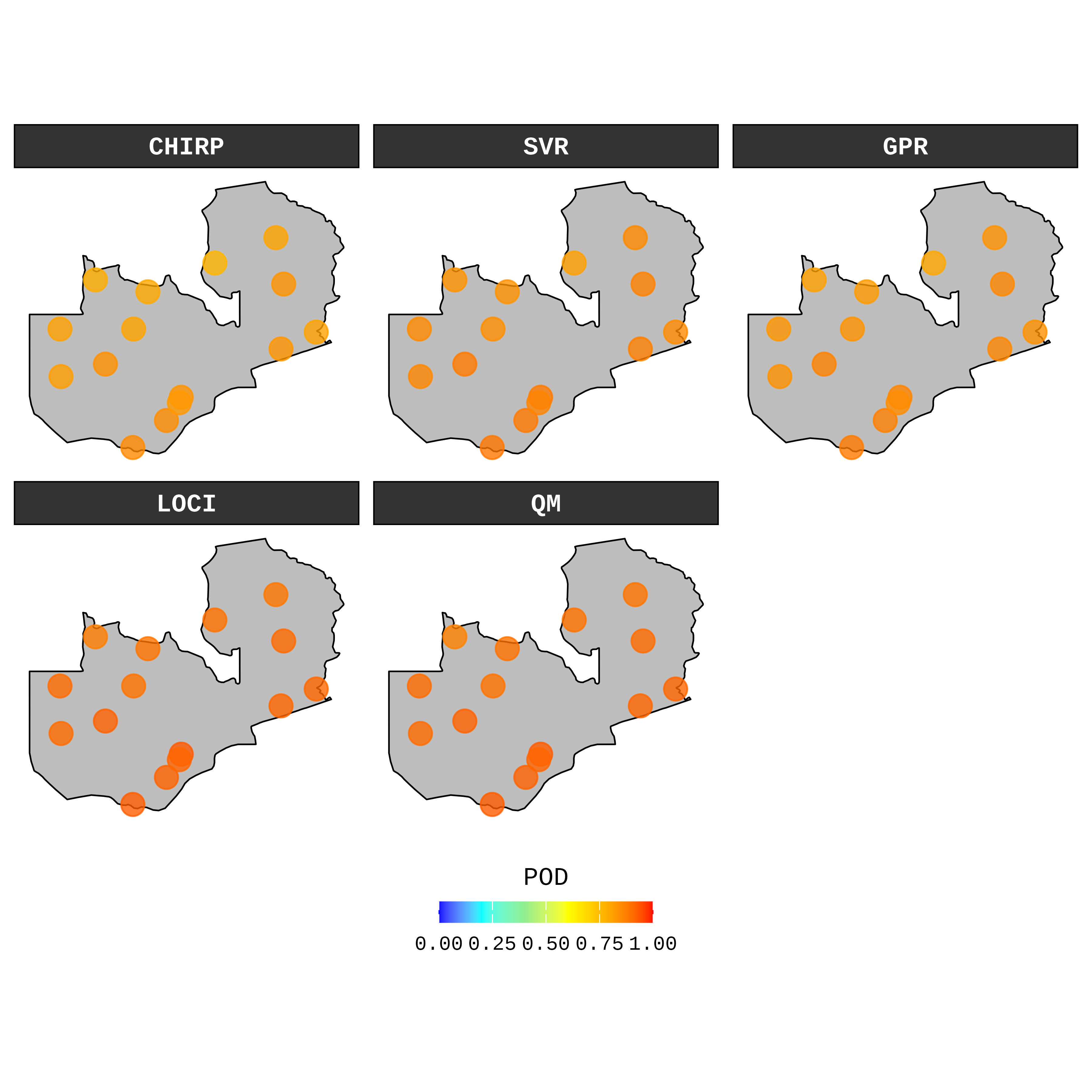}
	\caption{Probability of Detection (POD) of dry days by the BC methods on CHIRP across different stations in Zambia \label{pod_dry_chirp_zm_map}} 
\end{figure}

\begin{figure}[H]
	\centering
	\includegraphics[width=10.5 cm]{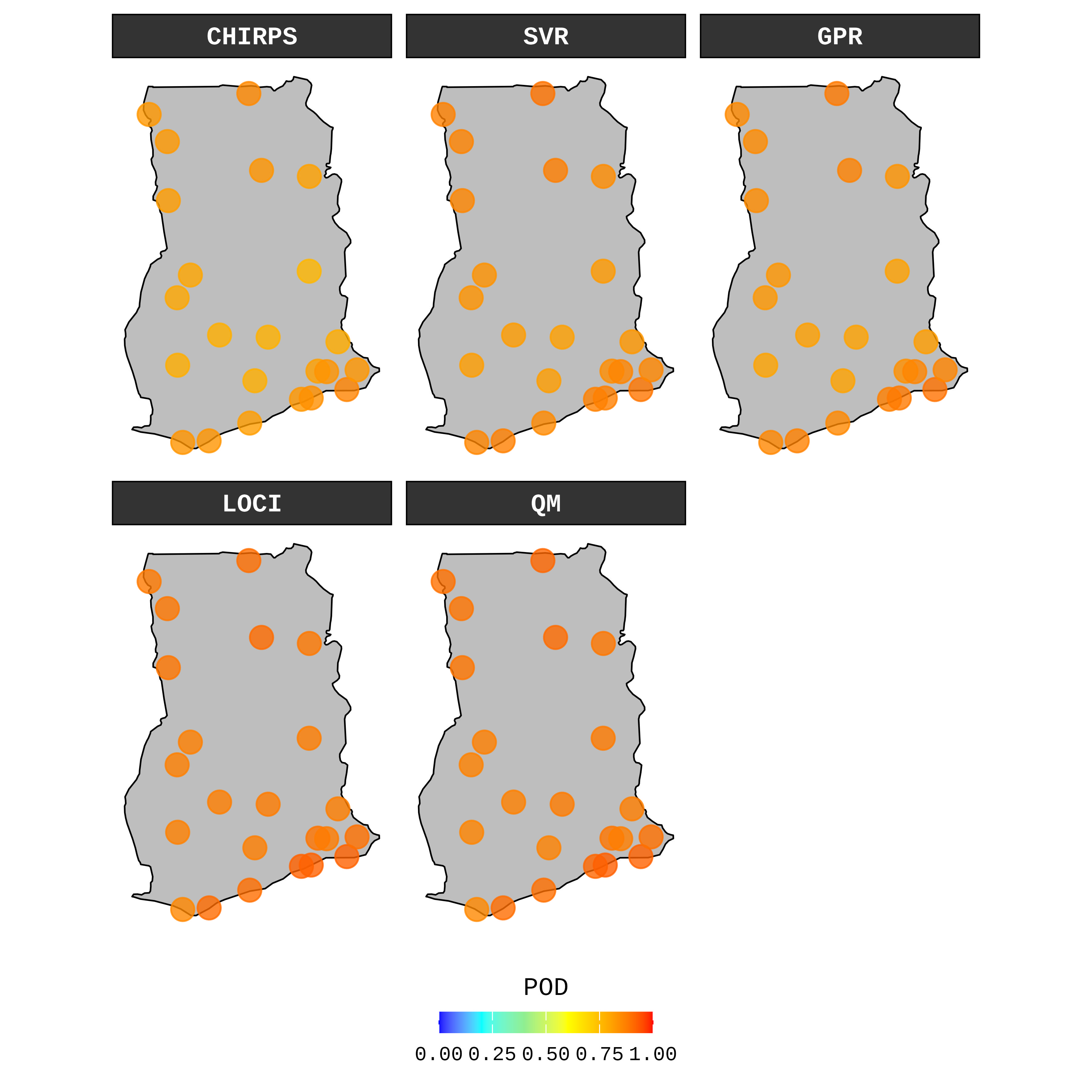}
	\caption{POD of dry days by the BC methods on CHIRPS across different stations in Ghana} \label{pod_dry_chirps_gh_map}
\end{figure}

All the SREs in their uncorrected form already showed high POD for dry days (POD $> 0.7$) across all stations in Zambia. This high detection rate can be seen in Figure~\ref{pod_dry_chirp_zm_map}, a map comparing the performance of the various BCs to uncorrected CHIRP, faceted by the methods. It was also the case for CHIRPS and TAMSAT in Ghana (CHIRPS shown in Figure~\ref{pod_dry_chirps_gh_map}). For these cases all the BC methods still increased the POD slightly further improving the detection of dry days. The improvement of the detection of dry days by LOCI and QM is not surprising since they specifically adjusted the number of rainy days. There is, however, a risk of missing some rainy days. For SVR and GPR which were targeted towards the estimation of intensities still managed to improve the detection of dry days. This is likely due to the inclusion of the additional predictors given them a bit of signal. 

\begin{figure}[H]
	\subfloat[]{
		\begin{minipage}[1\width]{
				0.5\textwidth}
			\centering
			\includegraphics[width=1\textwidth]{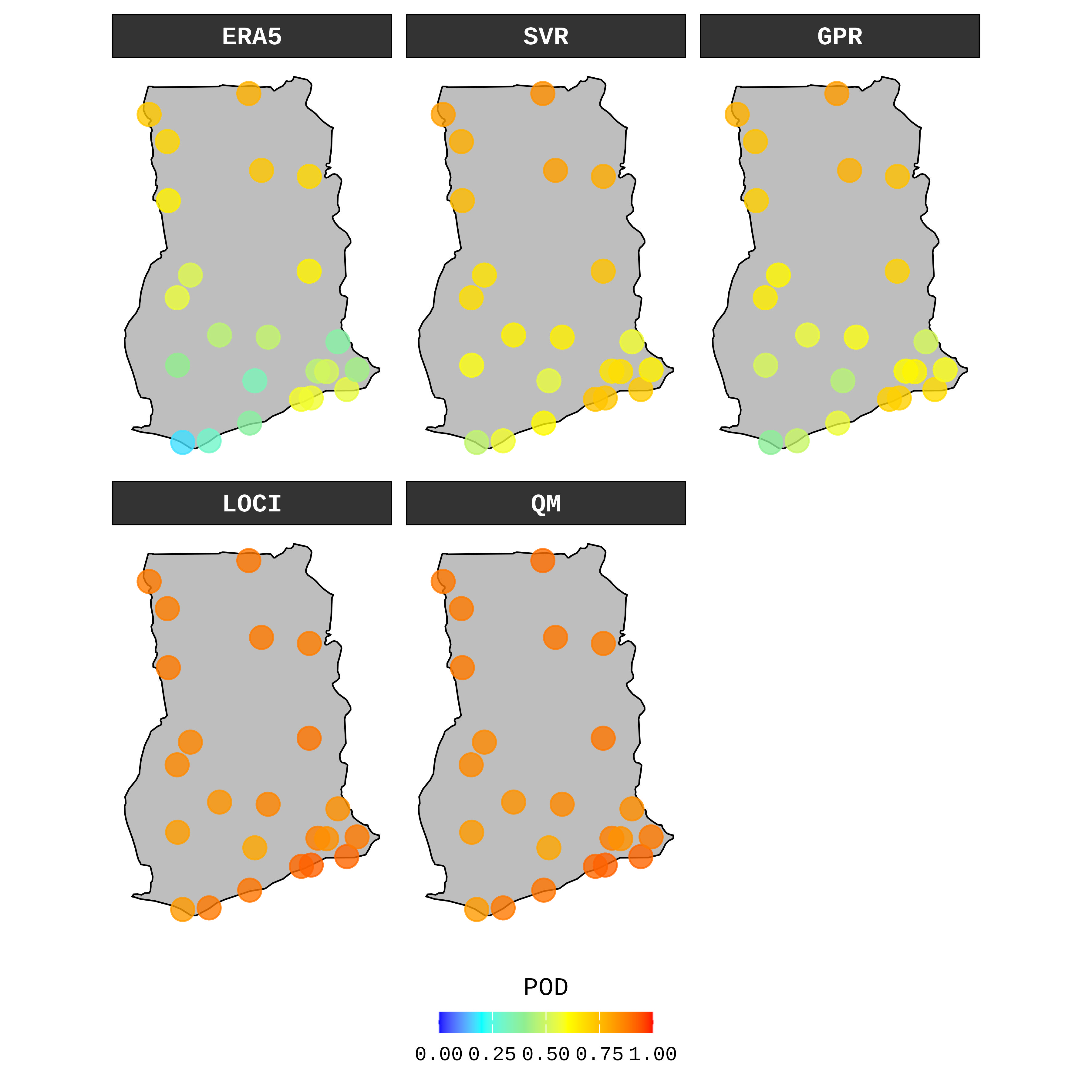}
	\end{minipage}}
	\hfill 	
	\subfloat[]{
		\begin{minipage}[1\width]{
				0.5\textwidth}
			\centering
			\includegraphics[width=1\textwidth]{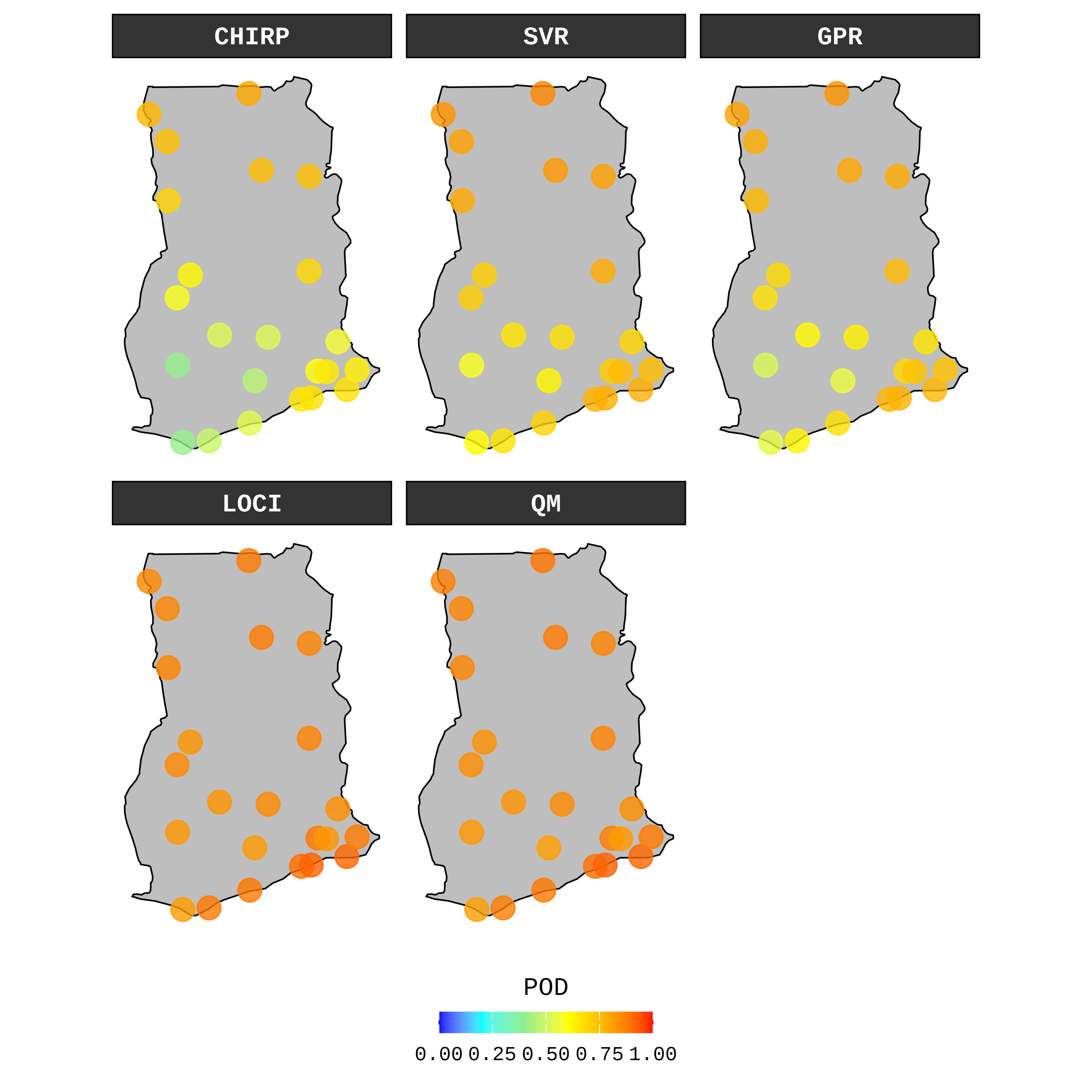}
	\end{minipage}}
	\caption{POD on the detection dry days by the BC methods when applied on ERA5 (a) and CHIRP (b) across different stations in Ghana}\label{pod_dry_era5_chirp_gh_map}
\end{figure}

For ERA5, AGERA5, PCDR, and CHIRP in Ghana, the detection of dry days by their uncorrected versions was quite poor in the Forest and Coastal zones (southen part of the country) while their performance in the Savannah zone (northern part) was comparable to the performance of CHIRPS and TAMSAT. Figures~\ref{pod_dry_era5_chirp_gh_map}(a) and \ref{pod_dry_era5_chirp_gh_map}(b) show the perfomance of the BC methods applied to ERA5 and CHIRP, respectively, compared to the uncorrected versions of these SREs for the detection of dry days across the stations in Ghana. As expected LOCI, QM slightly improved the detection of dry days in the Savannah zone, and significantly increased the POD in the Forest and Coastal zones. SVR and GPR slightly improved the detection of dry days across the country.

\subsubsection{Detection of heavy rains}\label{heavy_rains_detection}
\begin{figure}[H]
	\centering
	\includegraphics[width=10.5 cm]{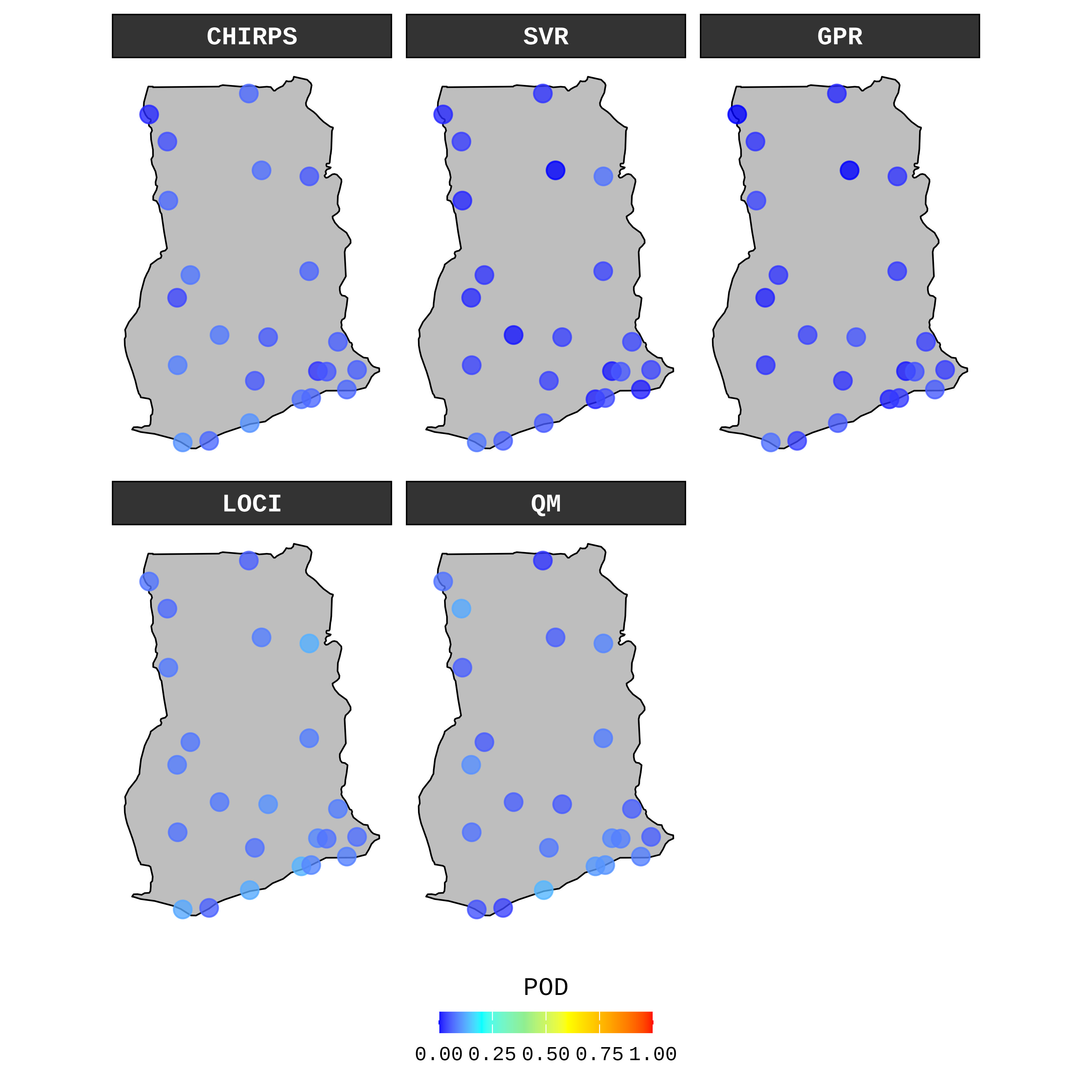}
	\caption{POD of heavy rains by the BC methods on CHIRPS across different stations in Ghana} \label{pod_heavy_chirps_gh_map}
\end{figure}

\begin{figure}[H]
	\centering
	\includegraphics[width=10.5 cm]{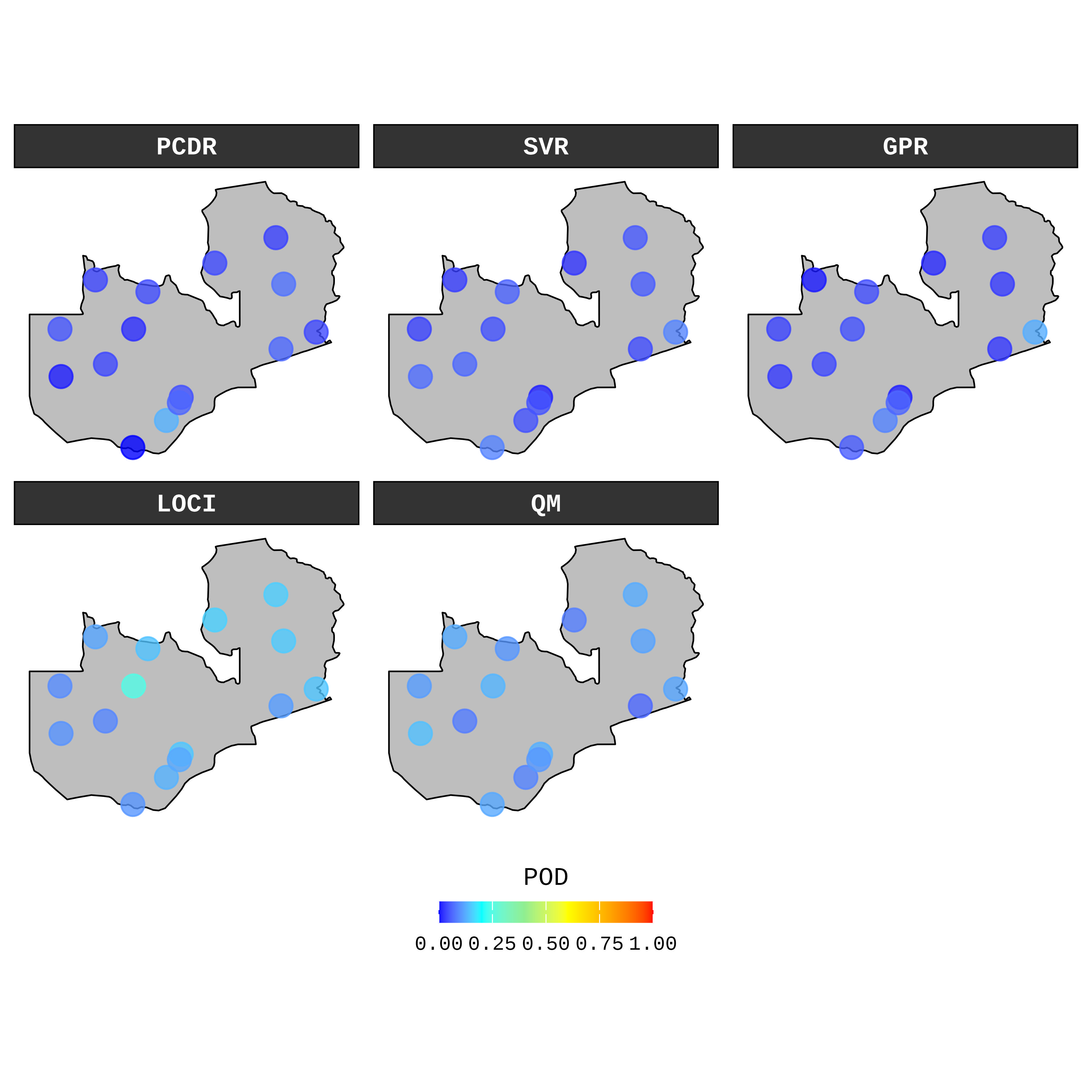}
	\caption{POD of heavy rains by the BC methods on PCDR across different stations in Zambia} \label{pod_heavy_pcdr_zm_map}
\end{figure}

All SREs in their uncorrected form demonstrated poor detection of heavy  rains $25 \leq \text{rain} < 40$) across Ghana and Zambia (with POD $< 0.2$) at most stations. SVR and GPR applied on the SREs (with the exception of ENACTS) tended to slightly worsen the detection of this rainfall event at most stations in both Ghana and Zambia, while LOCI and QM seemed to slightly improve it. This trend can be seen in Figures~\ref{pod_heavy_chirps_gh_map} and \ref{pod_heavy_pcdr_zm_map}. Figure~\ref{pod_heavy_chirps_gh_map} shows the performance of the BCs applied to CHIRPS compared to its uncorrected version across the stations in Ghana while Figure~\ref{pod_heavy_pcdr_zm_map} shows the performance of the BCs applied to PCDR compared to its uncorrected version across the stations in Zambia. The improvement by LOCI and QM were only slight increases in the POD, which generally still remained below 0.20 in both countries.

\unskip

\unskip

\begin{figure}[H]
	\centering
	\includegraphics[width=10.5 cm]{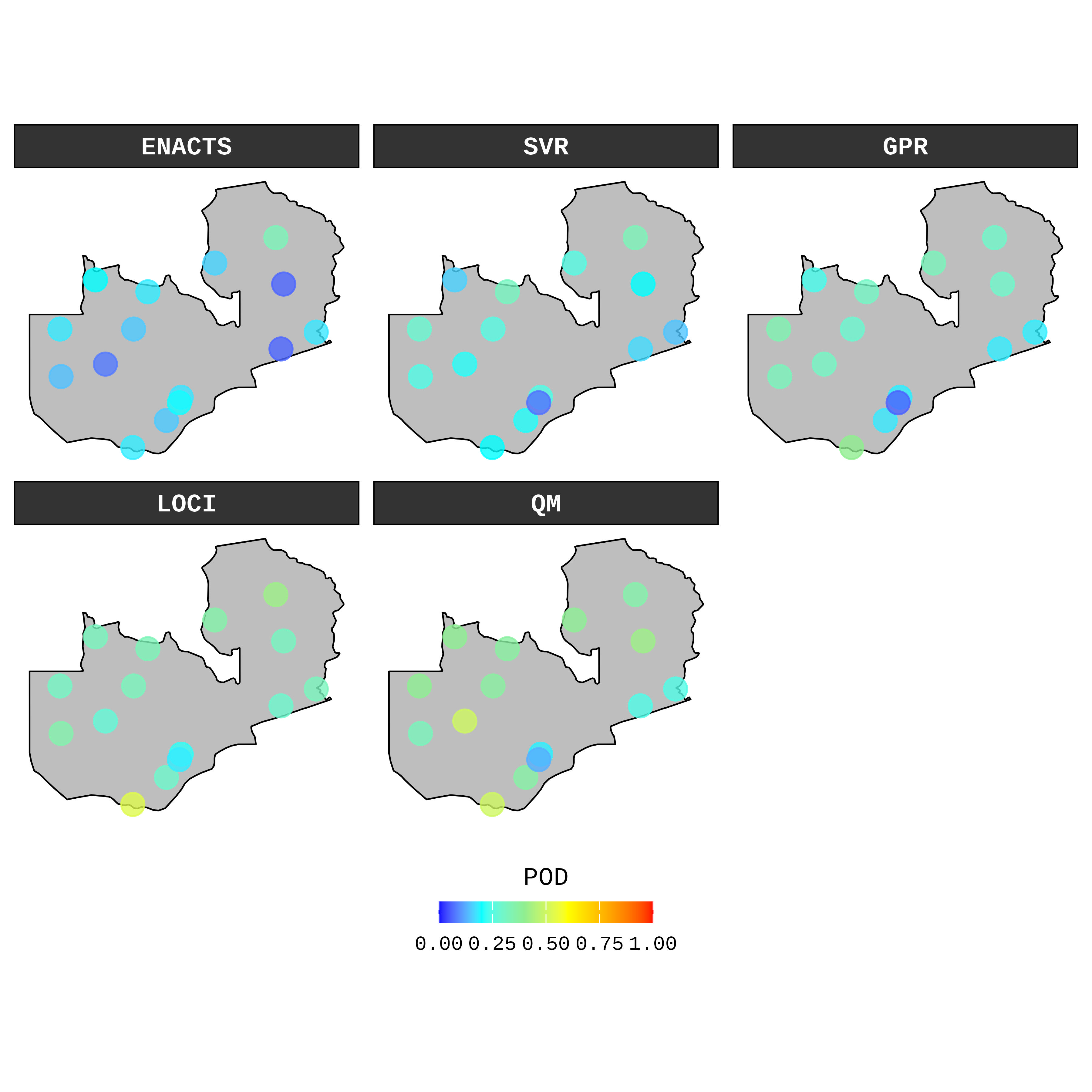}
	\caption{POD of heavy rains by the BC methods on ENACTS across different stations in Zambia} \label{pod_heavy_enacts_zm_map}
\end{figure}

ENACTS, on the other hand, showed quite a different behavior when the BCs were applied on it. All BCs seemed to increase the POD (see Figure~\ref{pod_heavy_enacts_zm_map}) at most stations indicating an improved capability to detect heavy rainfall in Zambia.

\unskip

Tables~\ref{tab:heavy_rain} presents the performance of the QM method on ENACTS for detecting heavy rainfall events in Zambia. Apart from Moorings,the QM- and LOCI-corrected ENACTS showed improved POD across all stations. Some stations experienced substantial increases, for example, Koama saw POD rise from 0.079 to 0.50 by QM. Moorings, however, had a slightly worsened POD by both LOCI and QM. It is worth noting that Moorings is the only independent station (not included in ENACTS). ENACTS performance at this stations suggests the product may not be that performant at independent stations. Similar enhancements were observed across Zambia for other BCs applied to ENACTS (as can be seen in Figure~\ref{pod_heavy_enacts_zm_map}). 

\begin{table}[H]
	\centering
	\caption{Performance comparison of ENACTS, QM-corrected ENACTS (QM), and LOCI for Heavy rain detection. Under Event Counts, the values under ENACTS, QM, and LOCI are the total model predictions (25-40 mm), combining true and false positives, while Observed is the total number of observed heavy rain events ($25 \leq \text{rain} < 40$)}
	\label{tab:heavy_rain}
	\begin{tabular}{lccccccc}
		\toprule
		\multirow{2}{*}{Station} & \multicolumn{3}{c}{POD} & \multicolumn{4}{c}{Event Counts} \\
		\cmidrule(lr){2-4} \cmidrule(lr){5-8}
		& ENACTS & QM & LOCI & Observed & ENACTS & QM & LOCI \\
		\midrule
		Chipata      & 0.178 & 0.244 & 0.319 & 135 & 98  & 149 & 171 \\
		Choma        & 0.153 & 0.361 & 0.292 & 72  & 51  & 87  & 77 \\
		Kaoma        & 0.079 & 0.500 & 0.272 & 114 & 47  & 115 & 88 \\
		Kasama       & 0.327 & 0.352 & 0.418 & 165 & 121 & 152 & 154 \\
		Kasempa      & 0.151 & 0.370 & 0.319 & 119 & 78  & 133 & 119 \\
		Livingstone  & 0.183 & 0.500 & 0.533 & 60  & 36  & 66  & 61 \\
		Magoye       & 0.175 & 0.186 & 0.216 & 97  & 58  & 82  & 75 \\
		Mansa        & 0.162 & 0.390 & 0.353 & 136 & 77  & 148 & 126 \\
		Mongu        & 0.144 & 0.322 & 0.347 & 118 & 58  & 106 & 106 \\
		Moorings     & 0.197 & 0.127 & 0.183 & 71  & 48  & 77  & 74 \\
		Mpika        & 0.061 & 0.417 & 0.313 & 115 & 37  & 128 & 100 \\
		Mwinilunga   & 0.200 & 0.393 & 0.317 & 145 & 91  & 145 & 119 \\
		Petauke      & 0.061 & 0.244 & 0.290 & 131 & 43  & 130 & 121 \\
		Solwezi      & 0.182 & 0.371 & 0.329 & 170 & 108 & 203 & 165 \\
		Zambezi      & 0.181 & 0.394 & 0.307 & 127 & 80  & 129 & 115 \\
		\bottomrule
	\end{tabular}
\end{table}

\subsubsection{Detection of violent rains}\label{violent_rains_detection}
\begin{figure}[H]
	\centering
	\includegraphics[width=10.5 cm]{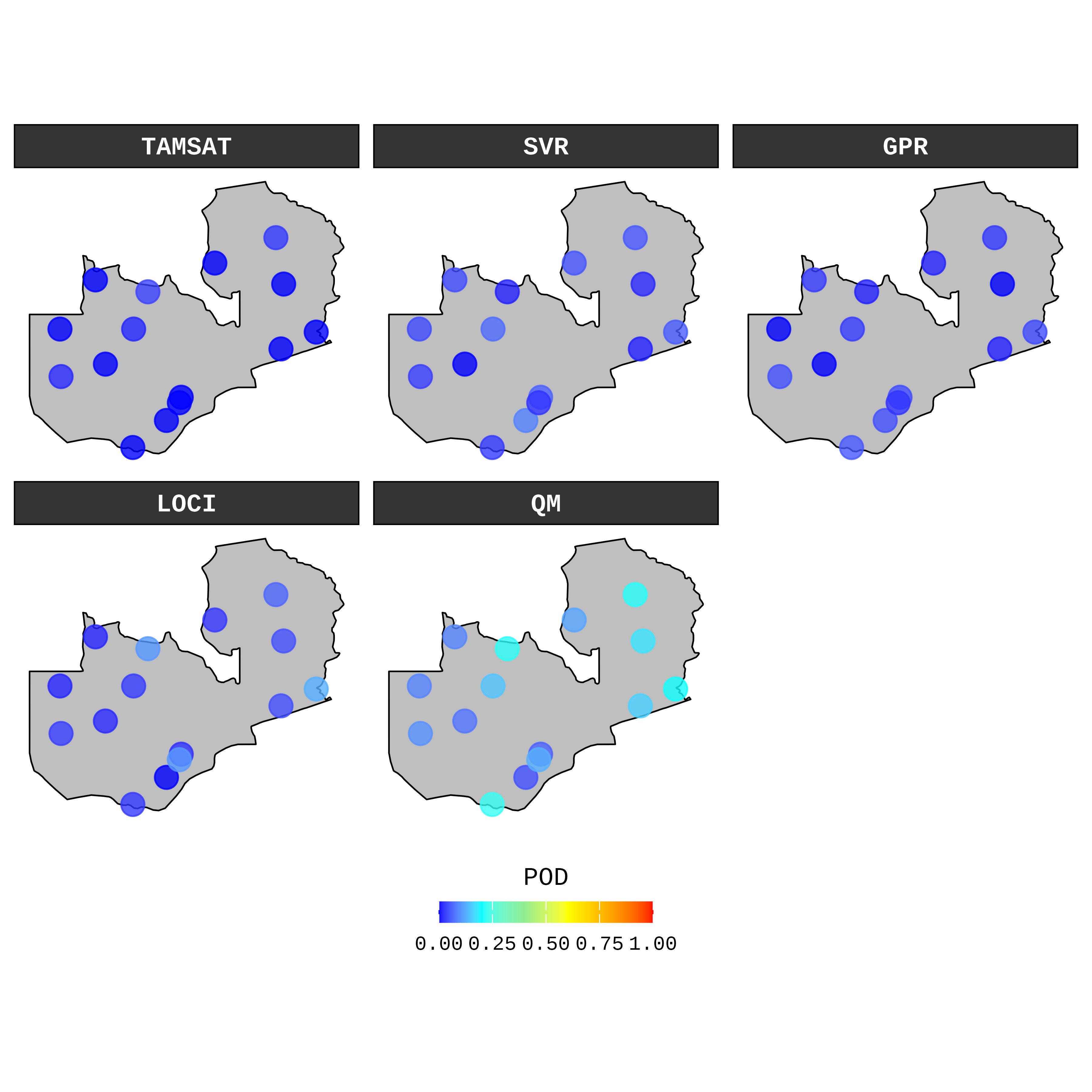}
	\caption{POD of violent rains by the BC methods on TAMSAT across different stations in Zambia} \label{pod_heavy_tamsat_zm_map}
\end{figure}

\begin{figure}[H]
	\centering
	\includegraphics[width=10.5 cm]{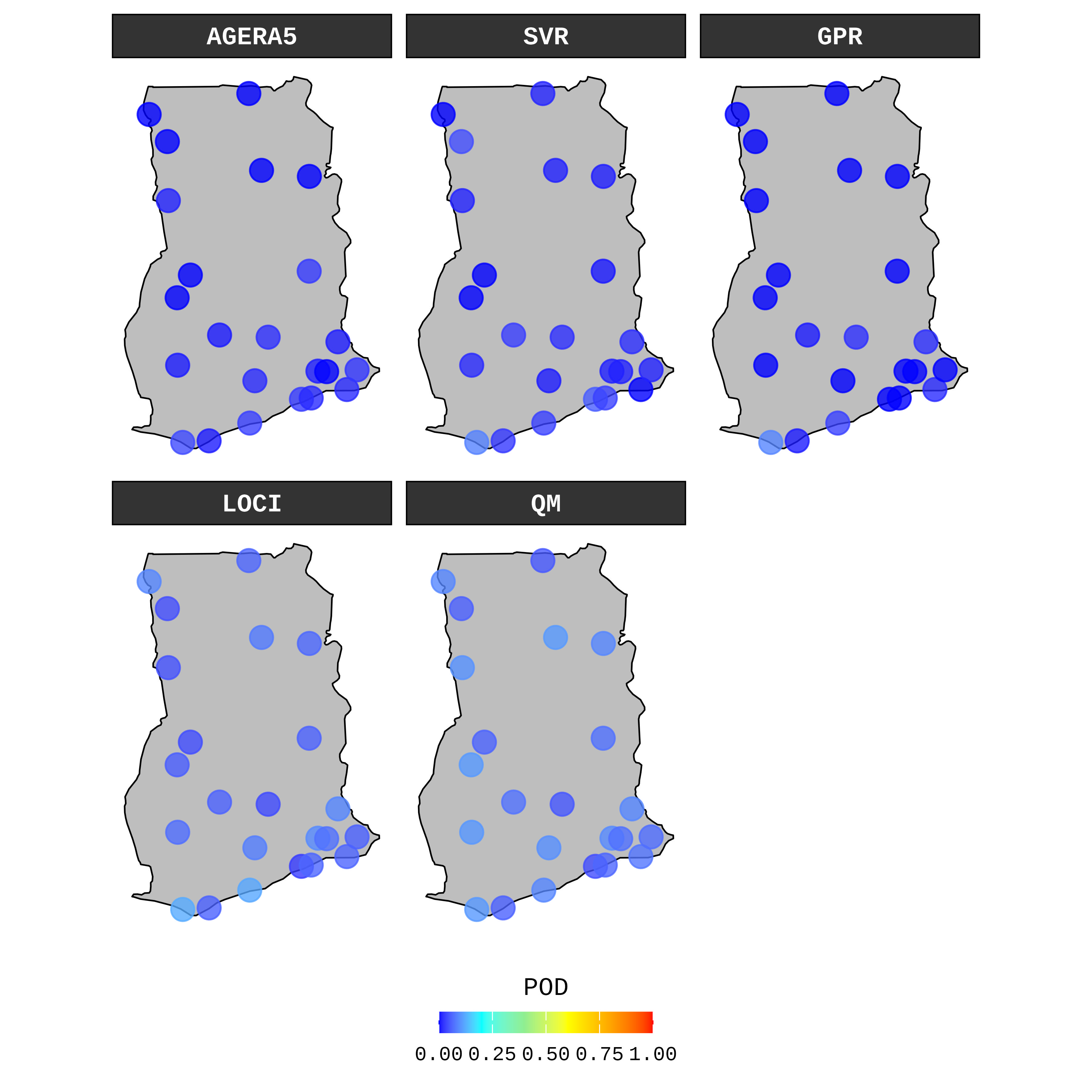}
	\caption{POD of violent rains by the BC methods on AGERA5 across different stations in Ghana} \label{pod_violent_agera5_gh_map}
\end{figure}

All the SREs in their uncorrected form had very low POD (close to $0$) for the detection of violent rains ($\ge$ 40 mm/day) at almost every station in Ghana and Zambia. The BCs applied on the SREs (except ENACTS) produced only slight increases or decreases in POD, which generally remained below $0.20$. LOCI and QM tended to slightly increase the POD in most cases for the SREs in Zambia, with QM showing the most increase (as can be seen in Figure~\ref{pod_heavy_tamsat_zm_map}, which shows the performance of the BCs applied to TAMSAT compared to its uncorrected version across the stations). Similar observations were made for the SREs in Ghana (see Figure~\ref{pod_violent_agera5_gh_map}, which shows the performance of the BC methods applied to AGERA5 compared to its uncorrected version across the stations). 
\unskip

\unskip

\begin{figure}[H]
	\centering
	\includegraphics[width=10.5 cm]{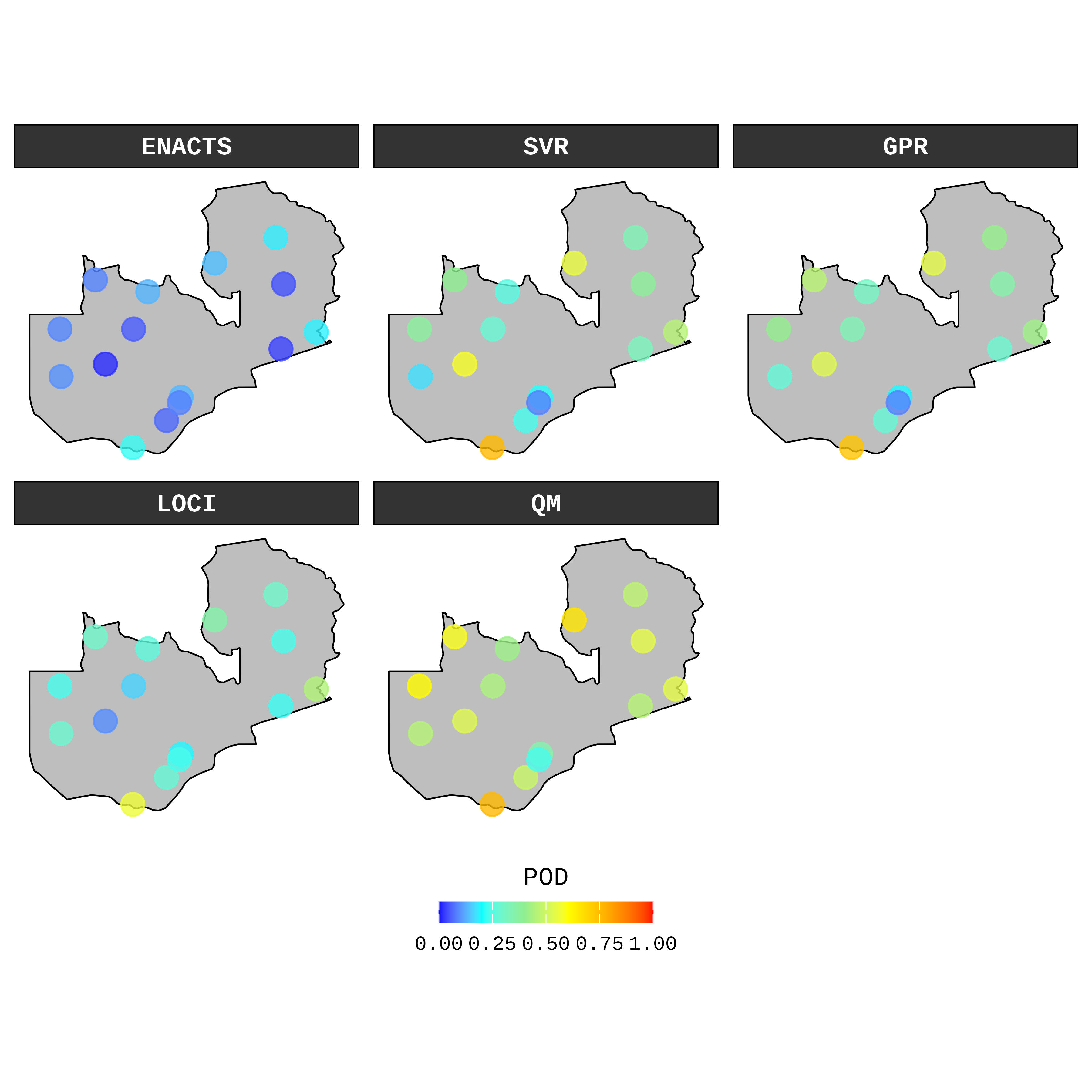}
	\caption{POD of violent rains by the BC methods on ENACTS across different stations in Zambia} \label{pod_violent_enacts_zm_map}
\end{figure}

All the BC methods applied on ENACTS seemed to improve the POD for the detection of violent rains (with POD > 0.2 at almost all stations). QM showed the most improvement (see Figure~\ref{pod_violent_enacts_zm_map}) at most stations.

\unskip

Table \ref{tab:violent_rain} compares the performance of the QM method against uncorrected ENACTS for detecting violent rainfall events in Zambia. The QM correction substantially improved the POD at most stations. Notably, at Chipata, Kaoma, Livingstone, Mansa and Mpika, the POD exceeded 0.50 after correction, a substantial increase from their initial low values.

\begin{table}[H]
	\centering
	\caption{Performance comparison of ENACTS, QM-corrected ENACTS (QM), and LOCI for Violent rain detection. Under Event Counts, the values under ENACTS, QM, and LOCI are the total model predictions ($\geq 40$), combining true and false positives, while Observed is the total actual violent rain events ($\text{rain} \geq 40$)}
	\label{tab:violent_rain} 
	\begin{tabular}{lccccccc}
		\toprule
		\multirow{2}{*}{Station} & \multicolumn{3}{c}{POD} & \multicolumn{4}{c}{Event Counts} \\
		\cmidrule(lr){2-4} \cmidrule(lr){5-8}
		& PRRAINT & QM & LOCI & Observed & ENACTS & QM & LOCI \\
		\midrule
		Chipata      & 0.188 & 0.541 & 0.451 & 133 & 27  & 131 & 95 \\
		Choma        & 0.064 & 0.489 & 0.277 & 47  & 5   & 46  & 21 \\
		Kaoma        & 0.016 & 0.532 & 0.097 & 62  & 2   & 41  & 28 \\
		Kasama       & 0.185 & 0.469 & 0.296 & 81  & 22  & 80  & 35 \\
		Kasempa      & 0.053 & 0.447 & 0.167 & 75  & 4   & 58  & 19 \\
		Livingstone  & 0.220 & 0.756 & 0.561 & 41  & 9   & 45  & 24 \\
		Magoye       & 0.135 & 0.346 & 0.192 & 52  & 12  & 51  & 25 \\
		Mansa        & 0.141 & 0.659 & 0.353 & 85  & 12  & 85  & 31 \\
		Mongu        & 0.170 & 0.457 & 0.286 & 70  & 16  & 69  & 39 \\
		Moorings     & 0.085 & 0.234 & 0.234 & 47  & 11  & 56  & 39 \\
		Mpika        & 0.044 & 0.544 & 0.235 & 68  & 3   & 59  & 19 \\
		Mwinilunga   & 0.090 & 0.590 & 0.295 & 78  & 7   & 69  & 27 \\
		Petauke      & 0.032 & 0.457 & 0.223 & 94  & 3   & 79  & 22 \\
		Solwezi      & 0.132 & 0.418 & 0.264 & 91  & 14  & 62  & 30 \\
		Zambezi      & 0.091 & 0.610 & 0.234 & 77  & 8   & 72  & 27 \\
		\bottomrule
	\end{tabular}
\end{table}

\subsection{Performance of the BC methods on seasonal scale}\label{seasonal_results} 

To model daily rainfall occurrence, a Zero-Order Markov chain approach was used. The probability of rainfall occurrence for each day of the year was derived using logistic regression, which incorporated Fourier series terms to account for cyclical seasonal patterns. The analysis was restricted to days where concurrent data were available from all sources (gauge observations, SREs, and BC outputs) to guarantee a consistent temporal comparison.

\begin{figure}[H]
	\centering
	\includegraphics[width=10.5 cm]{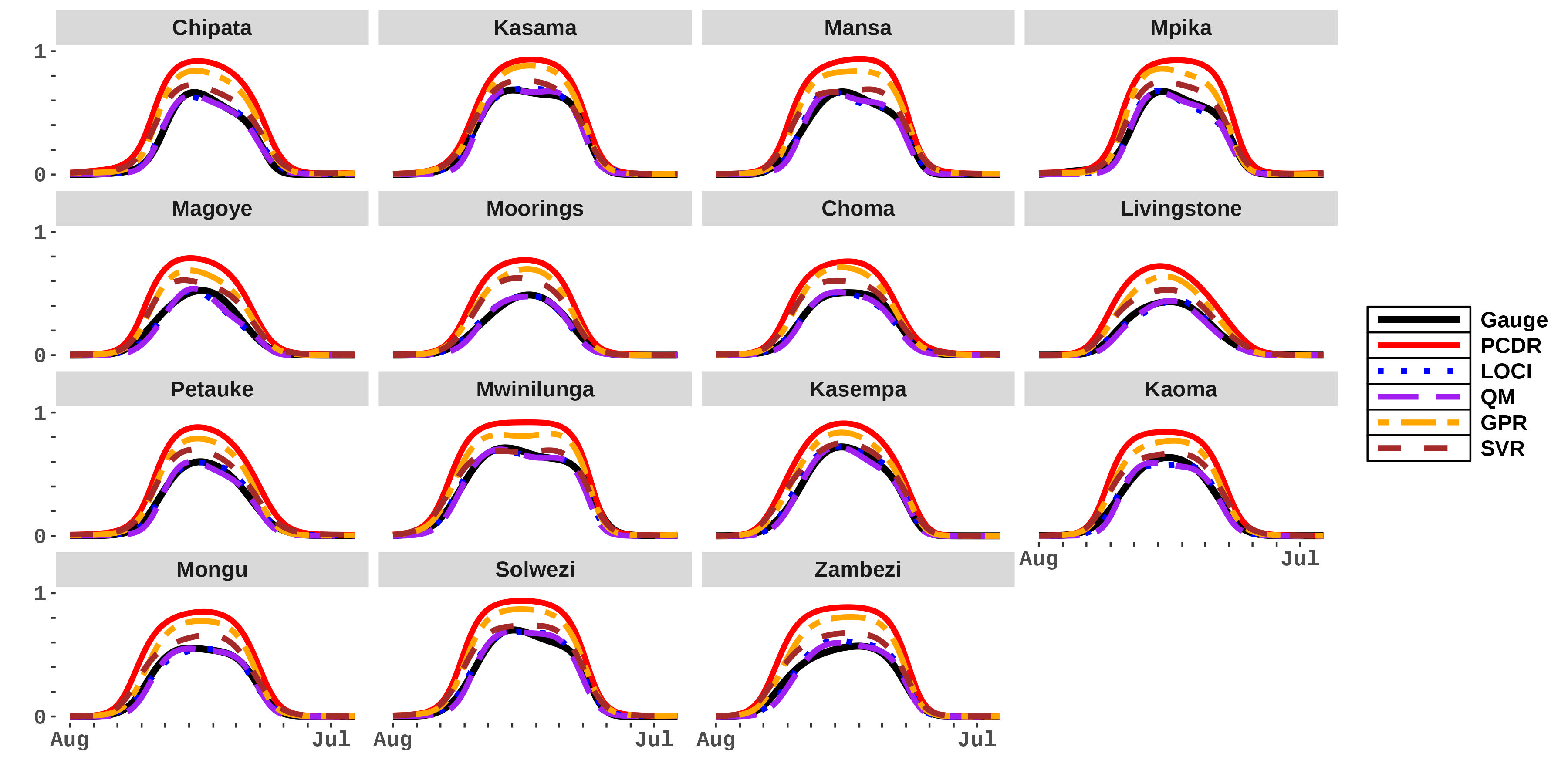}
	\caption{Performance of the BC methods on PCDR for capturing rainfall seasonality across the stations in Zambia. The y-axis represents the probability of rain, while the x-axis represents the day of year (starting August and ending July) \label{markov_rain_pcdr_zm}} 
\end{figure}

\begin{figure}[H]
	\centering
	\includegraphics[width=10.5 cm]{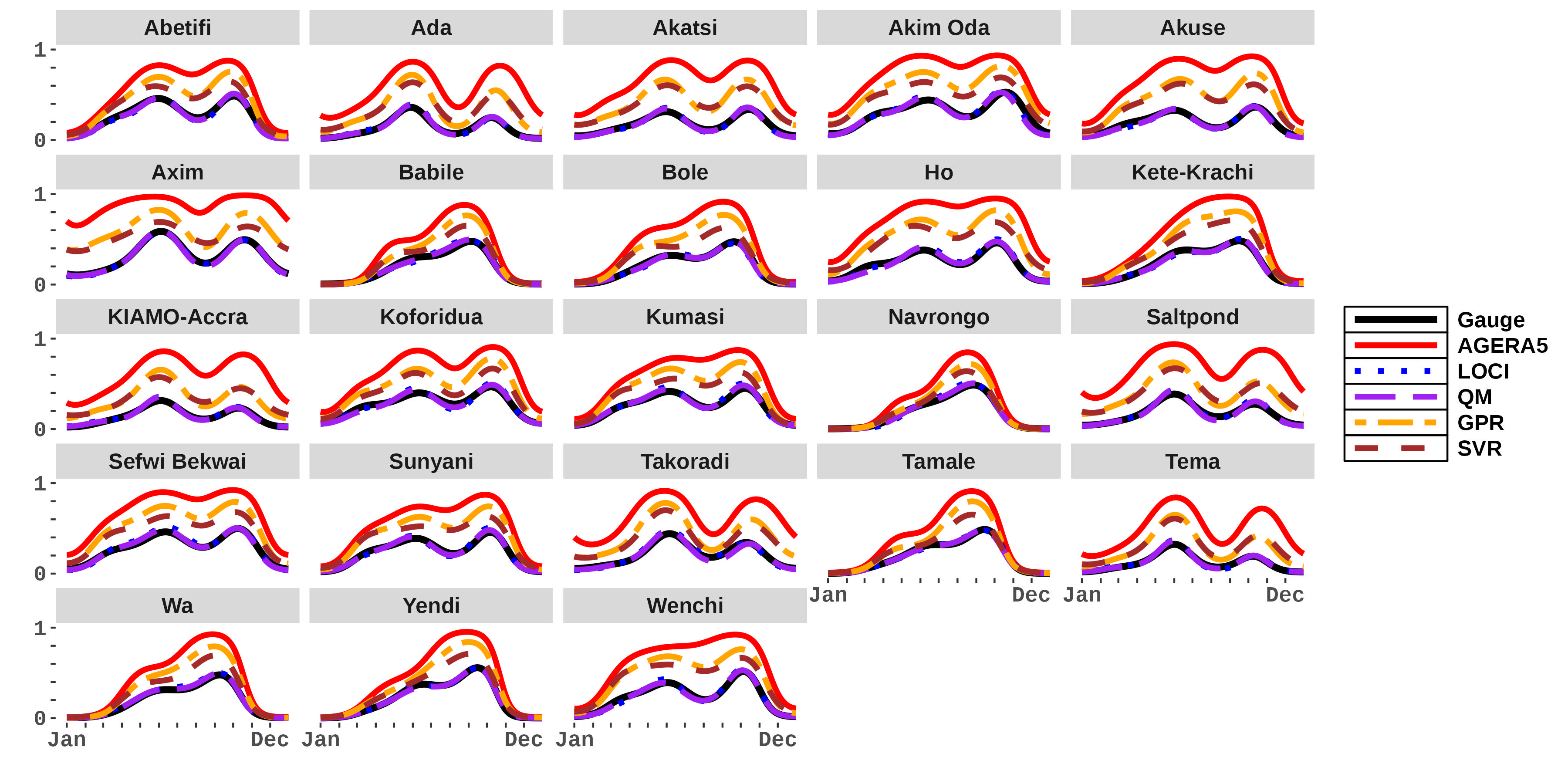}
	\caption{Performance of the BC methods on AGERA5 for capturing rainfall seasonality across the stations in Ghana. The y-axis represents the probability of rain, while the x-axis represents the day of year (January - December) \label{markov_rain_agera5_gh}} 
\end{figure}

\begin{figure}[H]
	\centering
	\includegraphics[width=10.5 cm]{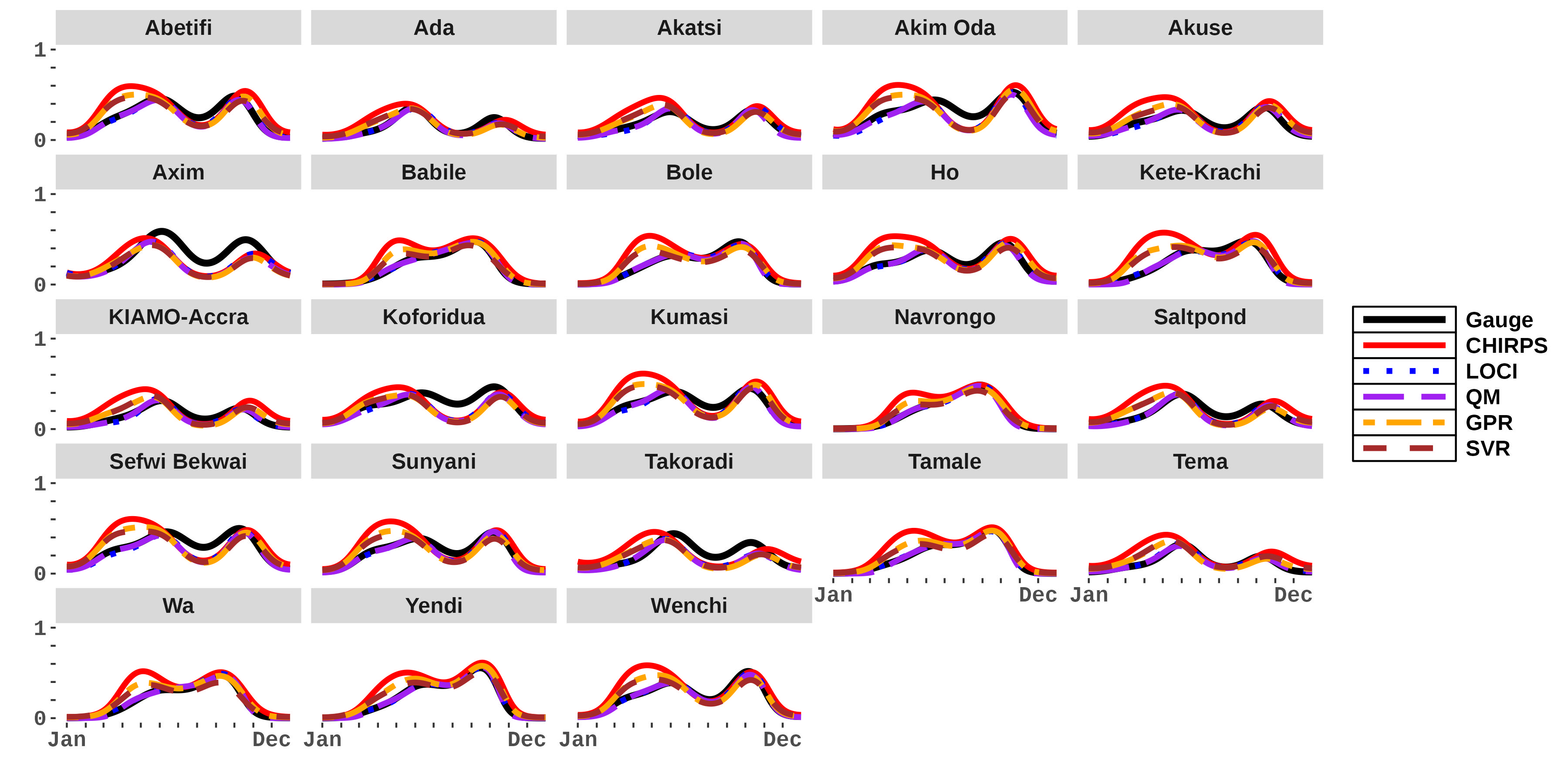}
	\caption{Performance of the BC methods on CHIRPS for capturing rainfall seasonality across the stations in Ghana. The y-axis represents the probability of rain, while the x-axis represents the day of year (January - December)} \label{markov_rain_chirps_gh} 
\end{figure}

\begin{figure}[H]
	\centering
	\includegraphics[width=10.5 cm]{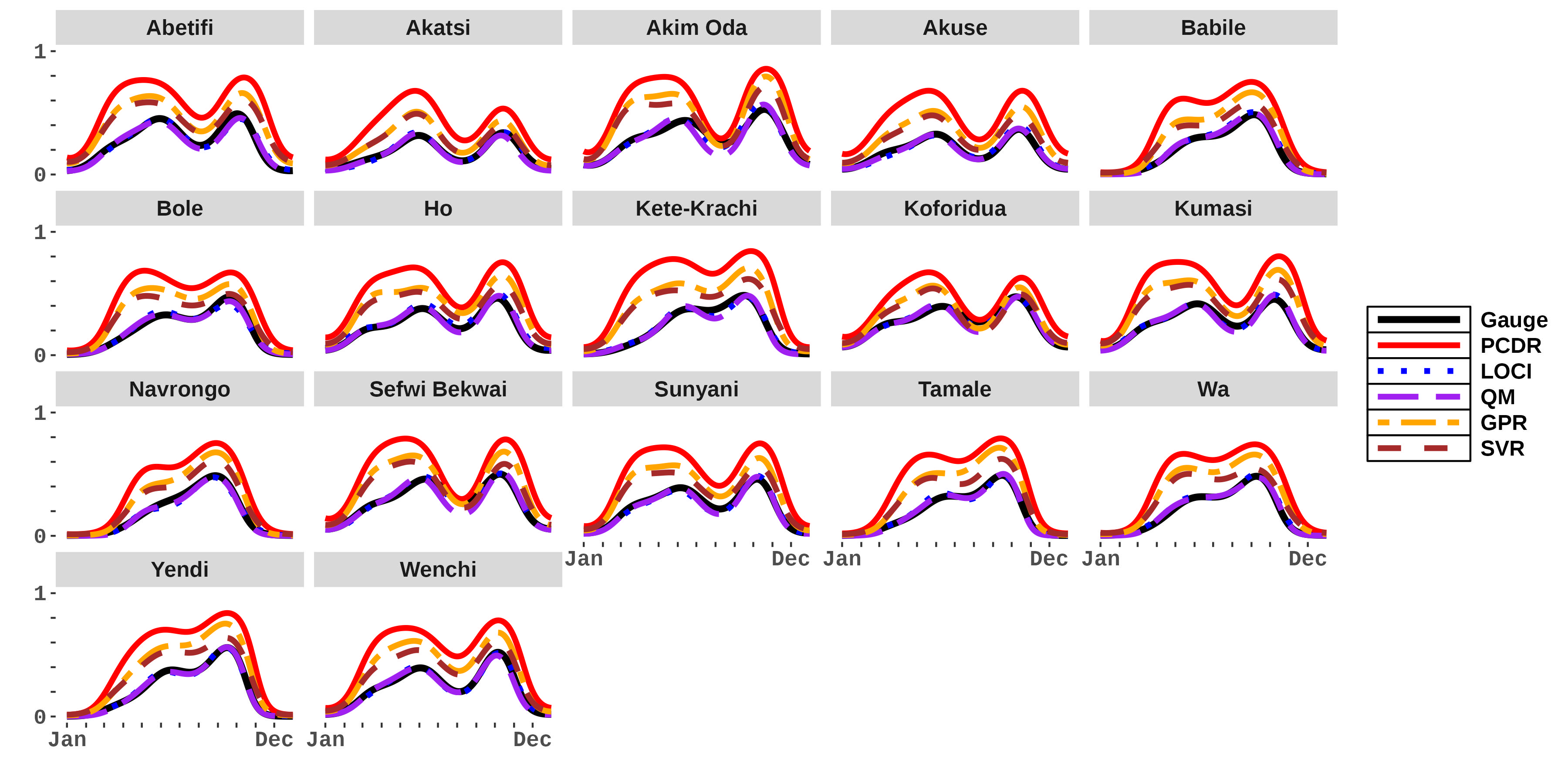}
	\caption{Performance of the BC methods on PCDR for capturing rainfall seasonality across the stations in Ghana. The y-axis represents the probability of rain, while the x-axis represents the day of year (January - December) \label{markov_rain_pcdr_gh}} 
\end{figure}

Figures~\ref{markov_rain_pcdr_zm} through~\ref{markov_rain_pcdr_gh} were generated using an identical methodology. In these figures, the observed rain day frequency from gauge data, using a 0.85 mm threshold, is represented by a solid black curve. The corresponding frequency from the SRE data is depicted by a solid red curve. The results from the various BC methods are illustrated by coloured dashed curves. The y-axis indicates the proportion of rain days, and the x-axis represents the date, ranging from August 1 to July 31 for Zambia and from January 1 to December 31 for Ghana.

QM- and LOCI-corrected versions captured the seasonality well at most stations (have similar dome shapes as the gauge curve in solid black line) both in Ghana and Zambia (Figures~\ref{markov_rain_pcdr_zm} \ref{markov_rain_agera5_gh}, \ref{markov_rain_chirps_gh}, and \ref{markov_rain_pcdr_gh}) and aligned with the gauge proportion of rainy days at most stations, especially for Zambia. This is expected as these methods do not specifically adjusted the rain day frequency.

SVR- and GPR-corrected SREs also captured the seasonal patterns well at the various. They also reduced the overestimations seen in the uncorrected SREs, where SVR sometimes align closly with the gauge curve (see Figure~\ref{markov_rain_pcdr_zm}). GPR had the least effect in reducing the overestimation in the rainy day proportion.

At some of the stations in Ghana, the rainfall regime (bi-modal or unimodal) did not seem to be captured by the uncorrected SREs. For example, Figure~\ref{markov_rain_agera5_gh} shows the performance of the BC methods on AGERA5 for capturing rainfall seasonality across the stations in Ghana. In this figure, the uncorrected AGERA5 did not clearly show a bimodal rainfall pattern at some of the stations in the Forest zone such as Kumasi and Wenchi as it is for the observed data and the other BC methods. However the corrected versions depict these bimodal patterns.  Also, Figure~\ref{markov_rain_pcdr_gh} shows the performance of the BC methods on PCDR for capturing rainfall seasonality across the stations in Ghana. The uncorrected PCDR as well as its SVR- and GPR-corrected versions did not seem to clearly depict a unimodal rainfall pattern at some of the stations in the Savanna zone such as Bole. 

Persistent underestimations were observed in CHIRPS rain day proportions at some coastal and southern stations, particularly between the dual peaks of the rainy seasons (Figure~\ref{markov_rain_chirps_gh}). These systematic biases remained uncorrected by all the BC methods. Similar underestimation patterns were evident in TAMSAT data for a couple of stations in the Forest and Coastal zones (see Figure S12 in supplementary material).

\subsection{Performance of the BC methods on annual scale}\label{annual_results}  
\subsubsection{Number of rainy days}\label{number_of_rain_days_reults} 

The overall mean errors (MEs) for the number of rainy days by SRE and BC method across all stations are presented for Zambia and Ghana in Tables~\ref{tab:mean_errors_zm} and \ref{tab:mean_errors_gh}, respectively. In both countries, all uncorrected SREs overestimated the annual count of rainy days, resulting in positive MEs. In Zambia, ENACTS had
the lowest bias ($+6.460$ days), followed by CHIRPS ($+11.558$ days). The remaining SREs exhibited more substantial overestimates, with MEs ranging from over 18 to 55 days (Table~\ref{tab:mean_errors_zm}).

As anticipated, the QM and LOCI methods exhibited superiority in reducing these biases, with LOCI showing a better performance. SVR and GPR, also reduced the mean error, with SVR demonstrating superior performance to GPR.

\begin{table}[H]
	\centering
	\caption{Overall Mean Errors on number of rainy days by SRE and BC method across the stations in Zambia}\label{tab:mean_errors_zm}
	\begin{tabular}{l *{5}{S[table-format=2.3]}}
		\toprule
		\textbf{SRE} & 
		\multicolumn{1}{c}{\textbf{Uncorrected}} & 
		\multicolumn{1}{c}{\textbf{LOCI}} & 
		\multicolumn{1}{c}{\textbf{QM}} & 
		\multicolumn{1}{c}{\textbf{GPR}} & 
		\multicolumn{1}{c}{\textbf{SVR}} \\
		\midrule
		AGERA5 &	54.903 & 0.267 & -3.710 & 33.742 & 23.364 \\
		CHIRP &	45.581 & -2.023 & -4.687 & 25.346 & 14.258 \\
		CHIRPS &	11.558 & 0.101 & -3.922 & 3.548 & -3.880 \\
		ENACTS &	6.460 &	-2.288 &	-7.023 &	2.279&	-2.502 \\
		ERA5	& 55.521 &	0.014 &	-3.576 &	34.502 &	22.387\\
		PCDR	& 48.395 &	0.522 &	-2.433 &	30.611 &	19.420 \\
		TAMSAT &	18.350 & -2.221	& -6.447 &	9.507 &	1.502 \\
		\bottomrule
	\end{tabular}
\end{table}

\begin{table}[H]
	\centering
	\caption{Overall Mean Errors on number of rainy days by SRE and BC method across the stations in Ghana}\label{tab:mean_errors_gh}
	\begin{tabular}{l *{5}{S[table-format=2.3]}}
		\toprule
		\textbf{SRE} & 
		\multicolumn{1}{c}{\textbf{Uncorrected}} & 
		\multicolumn{1}{c}{\textbf{LOCI}} & 
		\multicolumn{1}{c}{\textbf{QM}} & 
		\multicolumn{1}{c}{\textbf{GPR}} & 
		\multicolumn{1}{c}{\textbf{SVR}} \\
		\midrule
		AGERA5    & 130.132 &	1.021&	0.523&	71.551&	57.030\\
		CHIRP     & 95.439  &	-1.798  & -0.878  &	49.259  &	38.992 \\
		CHIRPS    & 23.005 & -7.765 & -6.670 &	4.604 &	1.190 \\
		ERA5      & 140.432&	1.468&	1.105&	76.267&	61.498 \\
		PCDR      & 88.821 &	-0.656&	-0.986&	50.108&	38.698\\
		TAMSAT    & 51.365&	-2.286&	-1.307&	28.032&	19.807 \\
		\bottomrule
	\end{tabular}
\end{table}

The overestimation by uncorrected SREs was more pronounced in Ghana. As shown in Table~\ref{tab:mean_errors_gh}, the SREs with the largest biases were ERA5 ($+140.432$ days) and AGERA5 ($+130.132$ days), whereas CHIRPS exhibited the smallest overestimation ($+23.005$ days). Both QM and LOCI drastically reduced this bias. In comparison, SVR and GPR showed less improvement.

The performance of uncorrected SREs at the station level displayed a north-south gradient across Ghana. With the exceptions of CHIRPS and TAMSAT, the SREs demonstrated a greater overestimation of rainy days in the southern Forest and Coastal Zones compared to the northern Savannah zone. For these SREs, both LOCI and QM successfully reduced the biases, bringing the estimated values into close alignment with the observed number of rainy days across all stations. The SVR and GPR methods also reduced this bias, with SVR outperforming GPR.

\begin{figure}[H]
	\subfloat[]{
		\begin{minipage}[1\width]{
				0.5\textwidth}
			\centering
			\includegraphics[width=1\textwidth]{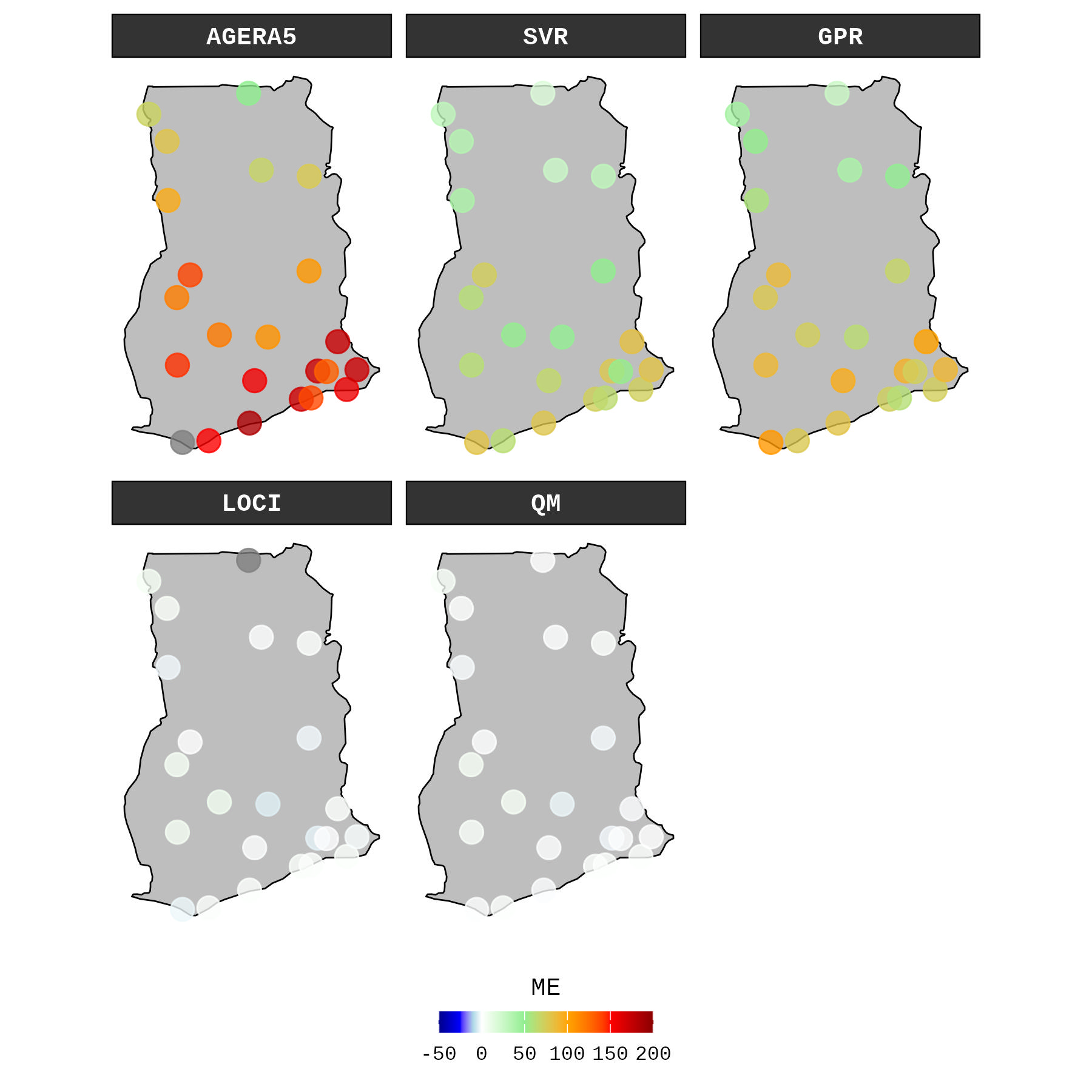}
	\end{minipage}}
	\hfill 	
	\subfloat[]{
		\begin{minipage}[1\width]{
				0.5\textwidth}
			\centering
			\includegraphics[width=1\textwidth]{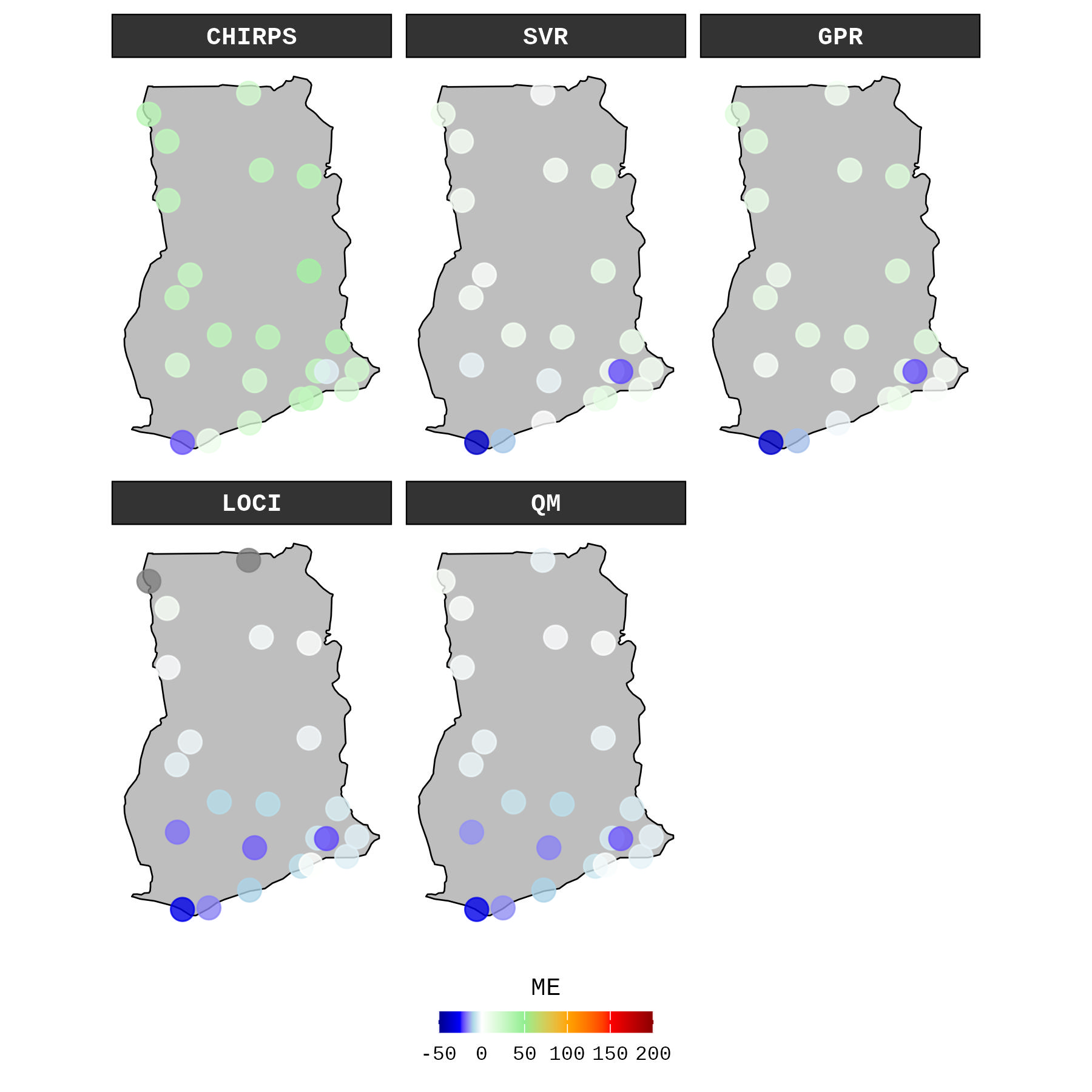}
	\end{minipage}}
	\caption{Mean Error (ME) on the annual number of rainy days across different stations in Ghana when the BC methods were applied on AGERA5 (a) and CHIRPS (b) }\label{n_days_agera5_chirps_gh}
\end{figure}

This spatial pattern is illustrated in Figure~\ref{n_days_agera5_chirps_gh}(a), which presents the MEs from the BC methods applied to AGERA5 for the annual number of rainy days at stations in Ghana. The figure confirms the north-south variation in overestimation for the uncorrected AGERA5 product. Both the LOCI- and QM-corrected versions of AGERA5 substantially reduced these biases while SVR and GPR also reduced across all stations.

In contrast, CHIRPS and TAMSAT showed a higher overestimation of the number of rainy days in the Savanna and Forest zones of the country. These overestimations were less pronounced in the far Coastal areas than in the northern locations in Ghana (Figure~\ref{n_days_agera5_chirps_gh}(b), and Figure~S4 respectively). All the BC methods reduced the biases, with LOCI- and QM-corrected values aligning closely with the observed number of rainy days at most stations in Ghana. However some underestimations were noted at some of the coastal and forest zones of the country (Figure~\ref{n_days_agera5_chirps_gh}(b) and Figure~S1).   

Some underestimations were noted in Zambia as well, although it was less pronounced in most cases (Figures~S1–S11).

These underestimations may be attributed to localized rainfall events that are not adequately captured by the models. Although the train-test split of the data was designed to minimize temporal discrepancies, the slight shift observed at some stations (Figures~\ref{fig_temp_split2} and \ref{fig_temp_split2_zm}) likely stems from factors such as localized precipitation events that remain unrepresented.

\subsubsection{Mean rain per rainy day}\label{annual_totals}
In this section the interest was to assess the performance of the BC methods on mean rain per rainy day, i.e, the total amount of rain in a year divided by the number of rainy days in the year. The BC methods was evaluated by their ability to reduce mean error (ME) in the SREs as well as improve variability. 

\begin{figure}[H]
	\subfloat[]{
		\begin{minipage}[1\width]{
				0.5\textwidth}
			\centering
			\includegraphics[width=1\textwidth]{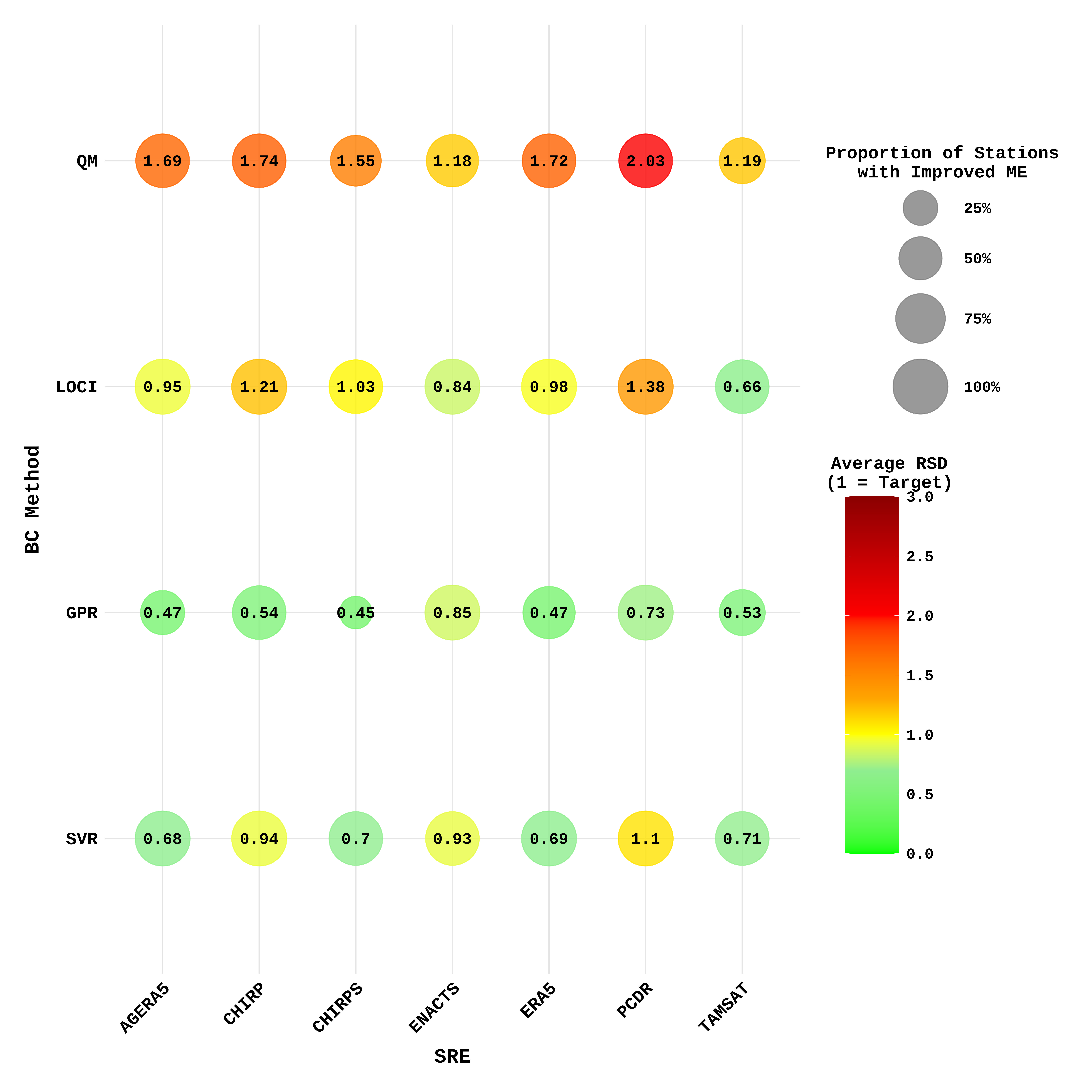}
	\end{minipage}}
	\hfill 	
	\subfloat[]{
		\begin{minipage}[1\width]{
				0.5\textwidth}
			\centering
			\includegraphics[width=1\textwidth]{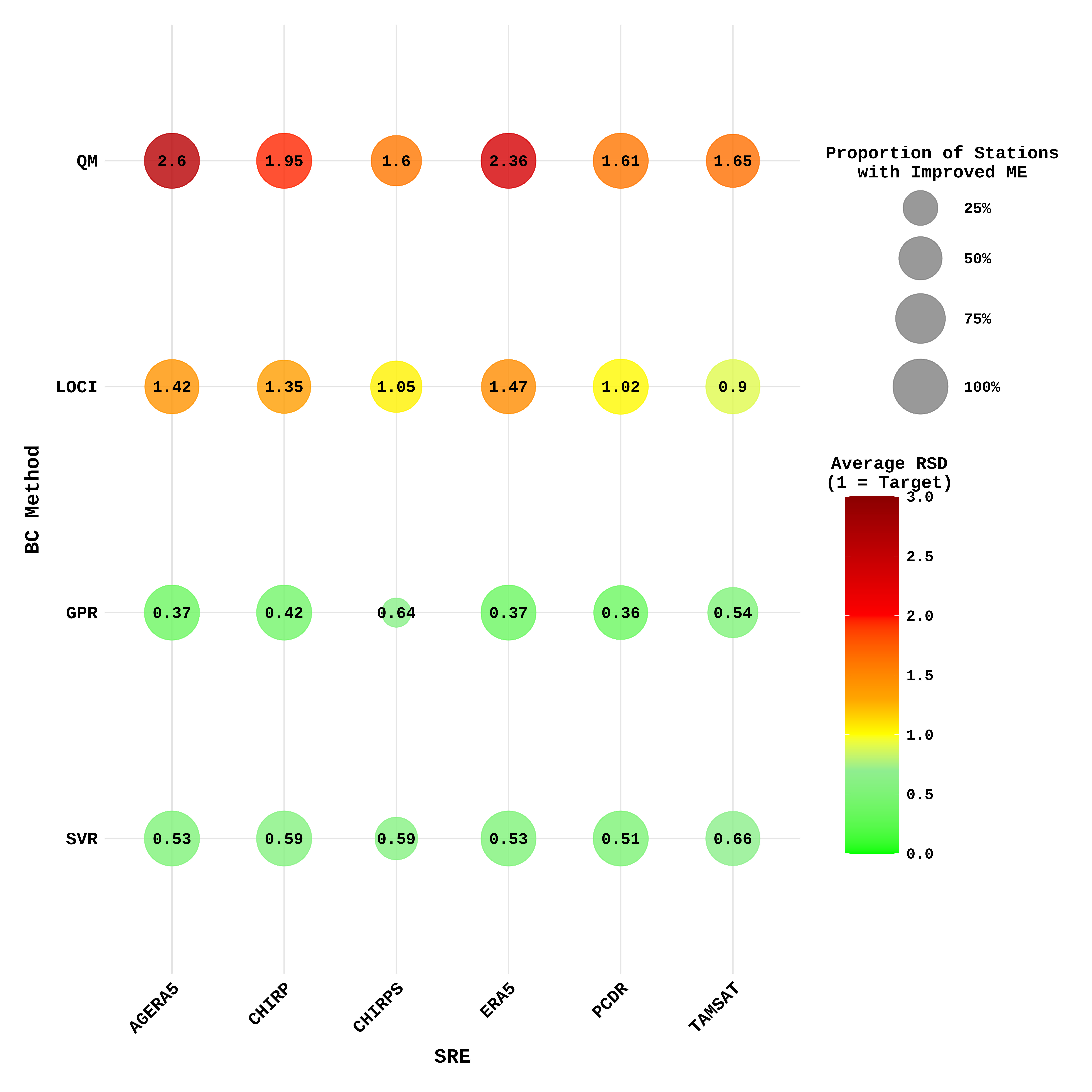}
	\end{minipage}}
	\caption{Bubble plots evaluating BC method performance in reducing ME on mean rain per rainy day for Zambia (a) and Ghana (b). Each bubble's position shows a BC method and SRE combination, its size shows the proportion of stations with reduced ME, and its color shows the mean RSD at those stations, also displayed inside the bubbles. Missing bubbles indicate no improvement}\label{bc_me<sre_me_mean_rain_zm_gh}
\end{figure}

Bubble plots in Figures~\ref{bc_me<sre_me_mean_rain_zm_gh}(a) and (b) display the proportion of stations in Zambia and Ghana, respectively, where a given BC method (y-axis) reduced the ME relative to the uncorrected SRE (x-axis). Bubble color represents the mean RSD across these stations, while bubble size corresponds to the proportion of stations. The mean RSD is also displayed numerically within each bubble. Missing bubbles indicate BC-SRE combinations that yielded no improvement.   

Based on Figure~\ref{bc_me<sre_me_mean_rain_zm_gh}(a), LOCI and QM successfully reduced the mean errors (MEs) for mean rain per rainy day at nearly all stations in Zambia across every SRE. This result is consistent with the performance of these BC methods, which systematically reduced the overestimation of rainy days found in most uncorrected SREs (Table~\ref{tab:mean_errors_zm}). Simultaneously, the corrected daily rainfall amounts remained within an acceptable range relative to the observed daily rainfall (Figure~\ref{bc_me_acceptable_zm_gh}(a)). LOCI typically produced a relative standard deviation (RSD) closer to 1 for almost all SREs, indicating variability similar to the observed mean rain per rainy day. In contrast, QM tended to overestimate variability across all SREs, likely due to the introduction of unobserved high values. Among the machine learning methods, SVR performed well, reducing bias at most stations and achieving an RSD near 1 at some. GPR also reduced bias at a large proportion of stations for most products, although it tended to be less variable than the observed rain day intensity. ENACTS was notable as the product for which all BC methods reduced the MEs. This is attributable to the ENACTS SREs having low MEs for daily rainfall amounts while also estimating a number of rainy days close to observations (Table~\ref{tab:mean_errors_zm}).

The trends in BC method performance for capturing mean rain per rainy day in Ghana (Figure~\ref{bc_me<sre_me_mean_rain_zm_gh}(b)) were, in most cases, similar to those observed in Zambia. The most notable difference lay in the magnitude of the relative mean RSD. In particular, QM exhibited a pronounced tendency to overestimate variability for all SREs, especially for AGERA5 and ERA5. Both SVR and GPR produced less variable values than the observed data.

\section{Discussion}\label{sec4}
This study provides new insights into the performance of a comprehensive suite of BC methods, including both statistical (LOCI, QM) and machine learning ML (SVR, GPR) approaches, applied to a wide range of SREs across the diverse climates of Ghana and Zambia. The analysis specifically evaluates how these BC methods affect the detection of rainfall events and the estimation of rainfall intensities.

ENACTS is generated by merging extensive ground station observations with bias-corrected CHIRPS or TAMSAT data~\cite{Dinku2022}. This contrasts with CHIRPS, which incorporates station data only on a five-day basis without full historical integration~\cite{Funk2015}, and TAMSAT, which uses station data purely for calibration rather than direct merging~\cite{Maidment2017}. ENACTS was the standout SRE in Zambia, with all BC methods performing well across multiple contexts. For example, nearly all the BC methods applied to ENACTS reduced the mean error (ME) at over 70\% of stations in Zambia (Figure \ref{bc_me<sre_me_zm_gh}a), even when the same methods had less effect in reducing errors for other SREs. The efficacy of BC methods on ENACTS is likely due to the comprehensive integration of station data within the product. For instance, nearly all BC methods applied to ENACTS enhanced the detection of heavy and violent rainfall events across almost all stations (e.g., Figure~\ref{pod_heavy_enacts_zm_map}, Table~\ref{tab:heavy_rain}), except at Moorings, where most BC methods decreased the probability of detection. Notably, Moorings is a station not included in the ENACTS. This indicates that the product’s performance may be less reliable at independent stations. Its performance at locations not included in this SRE therefore requires further investigation with more independent stations. The Zambia Meteorological Department is preparing data from volunteer stations not currently in ENACTS; whilst not yet available, these data could serve as a basis for a follow-up study.

The difficulty in accurately detecting heavy and violent rainfall events is a well-documented limitation of SREs~\cite{Yang2016, Mekonnen2023, ZambranoBigiarini2017, Ageet2022}. This limitation stems from the fundamental discrepancy between their areal estimates and point-based rain gauge measurements. Furthermore, recent evidence indicates that this challenge is resilient, often persisting even after BC techniques are applied~\cite{Li2023}. Our results corroborate this, indicating that, with the notable exception of ENACTS (which needs further investigation), the applied BC methods did not significantly improve SREs' ability to detect these events. The persistent low detection skill for heavy and violent rainfall (POD $\le 0.2$) across most BC methods has significant implications for flood-related applications. A low POD for these events would result in a failure to trigger flood warnings. Consequently, while the bias-corrected SREs evaluated here show improved mean performance, they should be used with extreme caution in flash-flood forecasting or urban drainage modeling. Future work should specifically target these extreme events. 
~\citet{Chen2010} demonstrated an approach where numerous meteorological variables were used to first develop a rain or no-rain classifier, followed by regression models applied exclusively to wet days. Building upon this, future research could focus on creating models specifically designed to identify heavy and violent rainfall days. Subsequently, regression models could be applied to these classified events to bias-correct their intensities. Existing research already shows promise in classifying such events. ~\citet{Yang2019} investigated the temporal clustering of heavy precipitation events in Europe using reanalysis datasets, revealing that these events indeed exhibit clustering patterns that reanalysis products effectively reproduce. The inclusion of diverse meteorological variables, similar to those in these studies, will likely provide strong signals for successfully classifying these critical events.

QM, along with other distribution-based techniques, has consistently demonstrated superior performance over other traditional statistical methods in various contexts of precipitation correction~\cite{cli12120226, LAKEW2020100741}. This advantage is largely attributed to QM’s ability to maintain the full statistical distribution of rainfall as well as the long term means, while simultaneously addressing both intensity and rain day frequency biases~\cite{Maraun2013}. Furthermore, some studies have shown that QM and other statistical methods can outperform ML approaches~\cite{DHAWAN2024e40352}. In line with these findings, the statistical methods in our study generally outperformed the ML approaches. However, a known limitation of QM is its tendency to inflate rainfall values~\cite{Maraun2013}, which was evident in our results, where QM-corrected SREs often overestimated rainfall amounts and increased variability relative to station observations. We also noted that the original LOCI has the potential to producing unrealistically high values for our locations. Our constrained version used here was mostly at par with QM, and usually produced estimates with similar variability to observed data than QM (e.g., Figure~\ref{bc_me<sre_me_mean_rain_zm_gh}).  

In contrast to our findings, ~\citet{Li2023}, who performed BC on IMERG-FR data over the Guangdong-Hong Kong-Macao Greater Bay Area, reported that ML methods (GPR and SVR) outperformed statical QM method. This discrepancy underscores that the performance of BC methods depends on various factors, including location, the SRE used, and regional characteristics. The length of data records may also be a factor; ~\citet{Li2023} acknowledged that the short historical record used to establish their transfer function might have limited QM's applicability during testing. However, the ML in our work have shown promise in reducing biases in rainfall intensities, with SVR showing superior performance to GPR. Even though these models are regression-based, they still managed to reduce the bias in rainy days (e.g., Tables~\ref{tab:mean_errors_zm} and \ref{tab:mean_errors_gh}).

The topography of southern Ghana (encompassing Forest and Coastal zones) is complex due to its vegetation, terrain, and proximity to the coast. These regions experience highly localized rainfall events~\cite{Amekudzi2015-qk}, which SREs struggle to detect accurately. We found that uncorrected SREs underestimated rainy days at some stations, a problem that was exacerbated by the BC methods, especially in Ghana's Forest and Coastal zones (see Figure~\ref{n_days_agera5_chirps_gh} and Supplementary Materials Figure~S4). These underestimations may be attributed to localized rainfall events that are not adequately captured by the models. Although the train-test split of the data in this work was designed to minimize temporal discrepancies, the persistent slight shift in rainy day thresholds observed at some stations (Figures~\ref{fig_temp_split2} and \ref{fig_temp_split2_zm}) likely stems from factors such as localized precipitation events that remain unrepresented. The more pronounced effect in southern Ghana may have resulted from noise caused by localized rainfall events that were captured differently between the training and test sets.

\section{Conclusion}\label{sec5}
This study rigorously evaluated a comprehensive suite of bias correction (BC) methods, including statistical approaches (LOCI, QM), and machine learning (SVR, GPR), applied to seven satellite rainfall estimates (SREs) across 38 stations in Ghana and Zambia, aiming to assess their performance in rainfall detection and intensity estimation. Notably, the LOCI method was implemented using a constrained version introduced in this work to prevent the generation of unrealistically high rainfall values observed with the original approach.

The study demonstrates that BC methods can significantly enhance the utility of SREs for hydrological applications in Ghana and Zambia. The ENACTS product, which uniquely integrates a large number of station records, emerged as the most amenable to corrections in Zambia, with nearly all BC methods successfully reducing mean error in daily rainfall amounts across most stations in Zambia (Figure \ref{bc_me<sre_me_zm_gh}a). However, its performance was not consistently high at Moorings, a station not incorporated into the ENACTS product. This indicates that the product’s performance may be less reliable at independent stations, highlighting the need for further validation at additional independent locations.

Statistical methods (QM and LOCI) generally outperformed machine learning approaches, though QM showed a tendency to inflate rainfall values.

All SREs corrected with QM and LOCI exhibited strong capability for detecting dry days (POD $\ge$ 0.80) across all stations. 


A critical limitation persists across most SREs and BC methods: the consistent failure to improve the detection of heavy and violent rainfall events (POD $\leq$ 0.2), with the exception of ENACTS. This shortcoming restricts their utility for applications such as flood risk assessment and highlights a crucial research gap. Future work should investigate the integration of additional predictor variables, focusing on the development of machine learning models specifically designed to identify these extreme rainfall days. Once classified, regression-based methods could then be applied to correct their intensities.

This work addresses key issues related to climate data accuracy in regions where agriculture is predominantly rain-fed. It provides a valuable reference for selecting appropriate BC methods and SREs for agricultural and other applications that depend on hydrological insights, such as regional hydrological management and climate adaptation strategies.

\section{Limitations of the Study}\label{sec6}
\begin{itemize}
	\item For the ML methods (SVR and GPR), an exhaustive grid search for optimal hyperparameters was infeasible due to the process being computationally intensive, involving a large number of potential models (a maximum of $\#$stations $\times$ $\#$SREs $\times$ $\#$categories). While using fixed hyperparameters might prevent the ML models from reaching peak optimization, these settings were selected through empirical trials to ensure stability and provide a consistent baseline for comparison across regions.
	
	\item Due to data scarcity (fewer than 10 training samples) within some specific rainfall intensity categories, a model was not trained for those cases, and the original SRE estimates were retained as the corrected data. Furthermore, no validation was performed within the training period due to this data scarcity. This limitation suggests that the comparative underperformance of ML methods is a robust finding rooted in the data regime. Simple statistical methods are fundamentally better suited for small dataset while complex models like GPR and SVR are data-hungry.
	
	\item The comparisons between SRE estimates and gauge observations in this study utilized a point-to-pixel approach. This inherent mismatch between the point measurement (gauge) and the area-averaged grid cell estimate (SRE pixel) can sometimes obscure the true performance of the SRE, which may have consequently affected the interpretation of our results.
	
	\item Performance metrics are provided as point estimates only. Point estimates omit the statistical uncertainty associated with small samples and extreme rainfall. Confidence intervals would distinguish significant performance differences from random variability, which is critical in data-sparse regions where outliers skew results. Uncertainty quantification would provide a more rigorous assessment of the stability and generalizability of the results.
	
	\item The performance metrics presented in this study are reported as point estimates without accompanying confidence intervals or measures of statistical uncertainty. While these estimates provide a clear, first-order comparison of bias correction methods across SREs and stations, they do not convey the variability or reliability associated with each metric. Incorporating uncertainty quantification would enhance the interpretability and robustness of the findings.
	
	\item While the bias correction methods used in this study are transferable to other regions, the specific findings presented here cannot not be generalized to other areas, even those with similar climatic regimes, without local validation.
	
	\item The evaluation of ENACTS is limited by a lack of independent validation stations. Because the product merges local station data, data circularity is a risk. With only the Moorings station available as an independent station, we are unable to confirm if superior performance reflects genuine bias correction or station reuse. Performance degradation at Moorings highlights the need for validation at more stations not used in the merging process.
\end{itemize}

\section*{Declarations}

\begin{itemize}
	\item \textbf{Funding} This publication was made possible by a grant from Carnegie Corporation of New York (provided through the African Institute for Mathematical Sciences). The statements made and views expressed are solely the responsibility of the author(s).
	
	\item \textbf{Competing interests} The authors have no relevant financial or non-financial interests to disclose.
	
	\item \textbf{Data Availability} The RE data used are all publicly available. The station data for Ghana and Zambia can be obtained from the Ghana Meteorological Agency, and the Zambia Meteorological Department, respectively. 
	
	\item \textbf{Code availability} Code is available upon a reasonable request.
	
	\item \textbf{Authors' contributions} 
	Conceptualization: J.B., D.S.; Methodology: J.B., D.S.; Formal analysis: J.B., D.S.; Data curation: J.B., D.S., F.F.T., D.P., S.O.A.; Writing --- original draft: J.B.; Writing --- review \& editing: J.B., D.S., D.N., F.F.T., D.P., S.O.A.; Visualization: J.B., D.S.; Supervision: D.S., D.N., F.F.T.
	
\end{itemize}

\bibliographystyle{elsarticle-harv}
\bibliography{cas-refs.bib}
\end{document}